\documentclass{amsart}
\usepackage{amsfonts}
\usepackage{amsmath,amscd}
\usepackage{amsthm}
\usepackage{amssymb}
\usepackage{latexsym}

\numberwithin{equation}{section}

\usepackage{color}

\newtheorem{rem}{Remark}
\newtheorem{notation}{Notation}

\numberwithin{equation}{section}
\begin{document}
\newcommand{\beqa}{\begin{eqnarray}}
\newcommand{\eeqa}{\end{eqnarray}}
\newcommand{\thmref}[1]{Theorem~\ref{#1}}
\newcommand{\secref}[1]{Sect.~\ref{#1}}
\newcommand{\lemref}[1]{Lemma~\ref{#1}}
\newcommand{\propref}[1]{Proposition~\ref{#1}}
\newcommand{\corref}[1]{Corollary~\ref{#1}}
\newcommand{\remref}[1]{Remark~\ref{#1}}
\newcommand{\er}[1]{(\ref{#1})}
\newcommand{\nc}{\newcommand}
\newcommand{\rnc}{\renewcommand}

\nc{\cal}{\mathcal}

\nc{\goth}{\mathfrak}
\rnc{\bold}{\mathbf}
\renewcommand{\frak}{\mathfrak}
\renewcommand{\Bbb}{\mathbb}

\newcommand{\id}{\text{id}}
\nc{\Cal}{\mathcal}
\nc{\Xp}[1]{X^+(#1)}
\nc{\Xm}[1]{X^-(#1)}
\nc{\on}{\operatorname}
\nc{\ch}{\mbox{ch}}
\nc{\Z}{{\bold Z}}
\nc{\J}{{\mathcal J}}
\nc{\C}{{\bold C}}
\nc{\Q}{{\bold Q}}
\nc{\oC}{{\widetilde{C}}}
\nc{\oc}{{\tilde{c}}}
\nc{\ocI}{ \overline{\cal I}}
\nc{\og}{{\tilde{\gamma}}}
\nc{\lC}{{\overline{C}}}
\nc{\lc}{{\overline{c}}}
\nc{\Rt}{{\tilde{R}}}

\nc{\odel}{{\overline{\delta}}}

\def\pr#1{\left(#1\right)_\infty}  

\renewcommand{\P}{{\mathcal P}}
\nc{\N}{{\Bbb N}}
\nc\beq{\begin{equation}}
\nc\enq{\end{equation}}
\nc\lan{\langle}
\nc\ran{\rangle}
\nc\bsl{\backslash}
\nc\mto{\mapsto}
\nc\lra{\leftrightarrow}
\nc\hra{\hookrightarrow}
\nc\sm{\smallmatrix}
\nc\esm{\endsmallmatrix}
\nc\sub{\subset}
\nc\ti{\tilde}
\nc\nl{\newline}
\nc\fra{\frac}
\nc\und{\underline}
\nc\ov{\overline}
\nc\ot{\otimes}

\nc\ochi{\overline{\chi}}
\nc\bbq{\bar{\bq}_l}
\nc\bcc{\thickfracwithdelims[]\thickness0}
\nc\ad{\text{\rm ad}}
\nc\Ad{\text{\rm Ad}}
\nc\Hom{\text{\rm Hom}}
\nc\End{\text{\rm End}}
\nc\Ind{\text{\rm Ind}}
\nc\Res{\text{\rm Res}}
\nc\Ker{\text{\rm Ker}}
\rnc\Im{\text{Im}}
\nc\sgn{\text{\rm sgn}}
\nc\tr{\text{\rm tr}}
\nc\Tr{\text{\rm Tr}}
\nc\supp{\text{\rm supp}}
\nc\card{\text{\rm card}}
\nc\bst{{}^\bigstar\!}
\nc\he{\heartsuit}
\nc\clu{\clubsuit}
\nc\spa{\spadesuit}
\nc\di{\diamond}
\nc\cW{\cal W}
\nc\cG{\cal G}
\nc\cZ{\cal Z}
\nc\ocW{\overline{\cal W}}
\nc\ocZ{\overline{\cal Z}}
\nc\al{\alpha}
\nc\bet{\beta}
\nc\ga{\gamma}
\nc\de{\delta}
\nc\ep{\epsilon}
\nc\io{\iota}
\nc\om{\omega}
\nc\si{\sigma}
\rnc\th{\theta}
\nc\ka{\kappa}
\nc\la{\lambda}
\nc\ze{\zeta}

\nc\vp{\varpi}
\nc\vt{\vartheta}
\nc\vr{\varrho}

\nc\odelta{\overline{\delta}}
\nc\Ga{\Gamma}
\nc\De{\Delta}
\nc\Om{\Omega}
\nc\Si{\Sigma}
\nc\Th{\Theta}
\nc\La{\Lambda}

\nc\boa{\bold a}
\nc\bob{\bold b}
\nc\boc{\bold c}
\nc\bod{\bold d}
\nc\boe{\bold e}
\nc\bof{\bold f}
\nc\bog{\bold g}
\nc\boh{\bold h}
\nc\boi{\bold i}
\nc\boj{\bold j}
\nc\bok{\bold k}
\nc\bol{\bold l}
\nc\bom{\bold m}
\nc\bon{\bold n}
\nc\boo{\bold o}
\nc\bop{\bold p}
\nc\boq{\bold q}
\nc\bor{\bold r}
\nc\bos{\bold s}
\nc\bou{\bold u}
\nc\bov{\bold v}
\nc\bow{\bold w}
\nc\boz{\bold z}

\nc\ba{\bold A}
\nc\bb{\bold B}
\nc\bc{\bold C}
\nc\bd{\bold D}
\nc\be{\bold E}
\nc\bg{\bold G}
\nc\bh{\bold H}
\nc\bi{\bold I}
\nc\bj{\bold J}
\nc\bk{\bold K}
\nc\bl{\bold L}
\nc\bm{\bold M}
\nc\bn{\bold N}
\nc\bo{\bold O}
\nc\bp{\bold P}
\nc\bq{\bold Q}
\nc\br{\bold R}
\nc\bs{\bold S}
\nc\bt{\bold T}
\nc\bu{\bold U}
\nc\bv{\bold V}
\nc\bw{\bold W}
\nc\bz{\bold Z}
\nc\bx{\bold X}

\nc\ca{\mathcal A}
\nc\cb{\mathcal B}
\nc\cc{\mathcal C}
\nc\cd{\mathcal D}
\nc\ce{\mathcal E}
\nc\cf{\mathcal F}
\nc\cg{\mathcal G}
\rnc\ch{\mathcal H}
\nc\ci{\mathcal I}
\nc\cj{\mathcal J}
\nc\ck{\mathcal K}
\nc\cl{\mathcal L}
\nc\cm{\mathcal M}
\nc\cn{\mathcal N}
\nc\co{\mathcal O}
\nc\cp{\mathcal P}
\nc\cq{\mathcal Q}
\nc\car{\mathcal R}
\nc\cs{\mathcal S}
\nc\ct{\mathcal T}
\nc\cu{\mathcal U}
\nc\cv{\mathcal V}
\nc\cz{\mathcal Z}
\nc\cx{\mathcal X}
\nc\cy{\mathcal Y}

\nc\e[1]{E_{#1}}
\nc\ei[1]{E_{\delta - \alpha_{#1}}}
\nc\esi[1]{E_{s \delta - \alpha_{#1}}}
\nc\eri[1]{E_{r \delta - \alpha_{#1}}}
\nc\ed[2][]{E_{#1 \delta,#2}}
\nc\ekd[1]{E_{k \delta,#1}}
\nc\emd[1]{E_{m \delta,#1}}
\nc\erd[1]{E_{r \delta,#1}}

\nc\ef[1]{F_{#1}}
\nc\efi[1]{F_{\delta - \alpha_{#1}}}
\nc\efsi[1]{F_{s \delta - \alpha_{#1}}}
\nc\efri[1]{F_{r \delta - \alpha_{#1}}}
\nc\efd[2][]{F_{#1 \delta,#2}}
\nc\efkd[1]{F_{k \delta,#1}}
\nc\efmd[1]{F_{m \delta,#1}}
\nc\efrd[1]{F_{r \delta,#1}}

\nc\fa{\frak a}
\nc\fb{\frak b}
\nc\fc{\frak c}
\nc\fd{\frak d}
\nc\fe{\frak e}
\nc\ff{\frak f}
\nc\fg{\frak g}
\nc\fh{\frak h}
\nc\fj{\frak j}
\nc\fk{\frak k}
\nc\fl{\frak l}
\nc\fm{\frak m}
\nc\fn{\frak n}
\nc\fo{\frak o}
\nc\fp{\frak p}
\nc\fq{\frak q}
\nc\fr{\frak r}
\nc\fs{\frak s}
\nc\ft{\frak t}
\nc\fu{\frak u}
\nc\fv{\frak v}
\nc\fz{\frak z}
\nc\fx{\frak x}
\nc\fy{\frak y}

\nc\fA{\frak A}
\nc\fB{\frak B}
\nc\fC{\frak C}
\nc\fD{\frak D}
\nc\fE{\frak E}
\nc\fF{\frak F}
\nc\fG{\frak G}
\nc\fH{\frak H}
\nc\fJ{\frak J}
\nc\fK{\frak K}
\nc\fL{\frak L}
\nc\fM{\frak M}
\nc\fN{\frak N}
\nc\fO{\frak O}
\nc\fP{\frak P}
\nc\fQ{\frak Q}
\nc\fR{\frak R}
\nc\fS{\frak S}
\nc\fT{\frak T}
\nc\fU{\frak U}
\nc\fV{\frak V}
\nc\fZ{\frak Z}
\nc\fX{\frak X}
\nc\fY{\frak Y}
\nc\tfi{\ti{\Phi}}
\nc\bF{\bold F}
\rnc\bol{\bold 1}

\nc\ua{\bold U_\A}

\nc\qinti[1]{[#1]_i}
\nc\q[1]{[#1]_q}
\nc\xpm[2]{E_{#2 \delta \pm \alpha_#1}}  
\nc\xmp[2]{E_{#2 \delta \mp \alpha_#1}}
\nc\xp[2]{E_{#2 \delta + \alpha_{#1}}}
\nc\xm[2]{E_{#2 \delta - \alpha_{#1}}}
\nc\hik{\ed{k}{i}}
\nc\hjl{\ed{l}{j}}
\nc\qcoeff[3]{\left[ \begin{smallmatrix} {#1}& \\ {#2}& \end{smallmatrix}
\negthickspace \right]_{#3}}
\nc\qi{q}
\nc\qj{q}

\nc\ufdm{{_\ca\bu}_{\rm fd}^{\le 0}}


\nc\isom{\cong} 

\nc{\pone}{{\Bbb C}{\Bbb P}^1}
\nc{\pa}{\partial}
\def\H{\mathcal H}
\def\L{\mathcal L}
\nc{\F}{{\mathcal F}}
\nc{\Sym}{{\goth S}}
\nc{\A}{{\mathcal A}}
\nc{\arr}{\rightarrow}
\nc{\larr}{\longrightarrow}

\nc{\ri}{\rangle}
\nc{\lef}{\langle}
\nc{\W}{{\mathcal W}}
\nc{\uqatwoatone}{{U_{q,1}}(\su)}
\nc{\uqtwo}{U_q(\goth{sl}_2)}
\nc{\dij}{\delta_{ij}}
\nc{\divei}{E_{\alpha_i}^{(n)}}
\nc{\divfi}{F_{\alpha_i}^{(n)}}
\nc{\Lzero}{\Lambda_0}
\nc{\Lone}{\Lambda_1}
\nc{\ve}{\varepsilon}
\nc{\bepsilon}{\bar{\epsilon}}
\nc{\bak}{\bar{k}}
\nc{\phioneminusi}{\Phi^{(1-i,i)}}
\nc{\phioneminusistar}{\Phi^{* (1-i,i)}}
\nc{\phii}{\Phi^{(i,1-i)}}
\nc{\Li}{\Lambda_i}
\nc{\Loneminusi}{\Lambda_{1-i}}
\nc{\vtimesz}{v_\ve \otimes z^m}

\nc{\asltwo}{\widehat{\goth{sl}_2}}
\nc\ag{\widehat{\goth{g}}}  
\nc\teb{\tilde E_\boc}
\nc\tebp{\tilde E_{\boc'}}

\newcommand{\LR}{\bar{R}}
\newcommand{\eeq}{\end{equation}}
\newcommand{\ben}{\begin{eqnarray}}
\newcommand{\een}{\end{eqnarray}}

\title[The half-infinite XXZ chain in Onsager's approach]{The half-infinite XXZ chain in Onsager's approach}
\author{P. Baseilhac}
\address{Laboratoire de Math\'ematiques et Physique Th\'eorique CNRS/UMR 7350,
     F\'ed\'eration Denis Poisson FR2964, Universit\'e de Tours, Parc de Grammont, 37200 Tours, FRANCE}
     \email{baseilha@lmpt.univ-tours.fr}
\author{S. Belliard}
\address{Laboratoire Charles Coulomb CNRS/UMR 5221, Universit\'e Montpellier 2, F-34095 Montpellier, FRANCE}
\email{samuel.belliard@univ-montp2.fr}
\begin{abstract} 
The half-infinite XXZ open spin chain with general integrable boundary conditions is considered within the recently developed `Onsager's approach'. Inspired by the finite size case, for any type of integrable boundary conditions it is shown that the transfer matrix is simply expressed in terms of the elements of a new type of current algebra recently introduced. In the massive regime $-1<q<0$, level one infinite dimensional representation ($q-$vertex operators) of the new current algebra are constructed in order to diagonalize the transfer matrix. For diagonal boundary conditions, known results of Jimbo {\it et al.} are recovered. For upper (or lower) non-diagonal boundary conditions, a solution is proposed. Vacuum and excited states are formulated within the representation theory of the current algebra using $q-$bosons, opening the way for the calculation of integral representations of correlation functions for a non-diagonal boundary.  Finally, for $q$ generic the long standing question of the hidden non-Abelian symmetry of the Hamiltonian is solved: it is either associated  with the $q-$Onsager algebra (generic non-diagonal case) or the augmented $q-$Onsager algebra (generic diagonal case). 
\end{abstract}


\maketitle

\vskip -0.6cm



\section{Introduction}
In the context of quantum integrable models, solutions of the planar Ising model in zero magnetic field have provided a considerable source of inspiration. In particular, among the non-perturbative approaches that have been considered in order to solve this model,  L. Onsager proposed in \cite{Ons44} to study the spectral problem for the transfer matrix using the representation theory of an infinite dimensional Lie algebra, the so-called Onsager algebra. Based on this approach, the largest and second largest eigenvalues of the transfer matrix of the model were obtained. Although the Onsager algebra was a central object in \cite{Ons44}, it received less attention in the following years than the star-triangle relations or the free fermion techniques - which didn't play any essential role in Onsager's original work - did. Despite of this, in the 1980s the Onsager algebra appeared to be closely related with the quantum integrable structure discovered by Dolan and Grady \cite{DG}. Then, it was understood that Hamiltonians of various integrable models \cite{PeAMT,Davies,Potts,Ar,Ahn} can be written solely in terms of the generators of the Onsager algebra acting on certain finite dimensional representations. For all these models, the integrability condition is encoded in a pair of relations, the so-called Dolan-Grady relations \cite{DG} - or, equivalently, in the defining relations of the Onsager algebra. In this formulation, all conserved quantities form an Abelian subalgebra of the Onsager algebra in correspondance with the Dolan-Grady hierarchy. As a consequence, using the explicit relation between the Onsager algebra and the loop algebra of $sl_2$ \cite{Davies}, a generic and rather simple formula for the spectrum of the Hamiltonian associated with any of these models was explicitly obtained. For many years, the range of applications of the approach initiated in \cite{Ons44,PeAMT,Davies,Potts,Ar,Ahn} - here named as the Onsager's approach - remained however limited to a subset of integrable models, and other consequences of the existence of the Onsager algebra were not further explored. \vspace{1mm}

This situation started to change in recent years, when  a $q-$deformed analog of the Onsager algebra was discovered as the integrability condition of a large class of quantum integrable models defined either on the lattice or in the continuum \cite{Bas1,Bas2,BK}: this algebraic structure was identified by considering in details the Sklyanin's operator that appears in the standard formulation of models with boundaries\footnote{This does not imply that the approach solely applies to models with boundaries: for $q=1$, the Ising \cite{Ons44} and superintegrable chiral Potts models \cite{PeAMT,Davies,Potts} are explicit counterexamples of this idea (see also \cite{Ar,Ahn}).} \cite{Skly88}, establishing for the first time a correspondance between the $q-$Onsager algebra and the reflection equation. Remarkably, for this class of  models the integrability condition consists in a pair of $q-$Dolan-Grady relations, or, alternatively, in the existence of an infinite dimensional $q-$deformed analog of the Onsager algebra proposed in \cite{BK}. Furthermore, all mutually commuting conserved quantities - for instance the Hamiltonian - generates an Abelian subalgebra called the $q-$Dolan-Grady hierarchy. Based on these results, a new interest for the Onsager's approach grew up. Inspired by Onsager's strategy for the solution of the two-dimensional Ising model \cite{Ons44} and later works on the superintegrable chiral Potts model \cite{Potts,PeAMT,Davies}, as well as the conformal field theory program \cite{BPZ}, it became clear  that finding a solution of a specific model which integrability condition is associated with the $q-$Onsager algebra could be considered through a detailed analysis of the finite or infinite dimensional representations involved. Knowing the essential problems arising within the algebraic Bethe ansatz framework  (see \cite{bXXZ} for more details) when applied to lattice models with integrable generic boundary conditions, the development of an alternative approach such that the Onsager's one became highly desirable.
\vspace{1mm}

 Up to now, the application of the Onsager's approach to lattice models\footnote{In the continuum, the existence of a hidden non-Abelian symmetry associated with a generalized $q-$Onsager algebra plays a central role in the derivation of scattering amplitudes associated with affine Toda field theories with a dynamical boundary, see \cite{BK0,BF}.} for $q\neq 1$ has been essentially restricted to the study of the {\it finite} XXZ open spin chain with {\it generic} integrable boundary conditions, in which case known results\footnote{For instance, the linear relations between the left and right boundary parameters that arise in the Bethe ansatz approach in order to construct a suitable reference state or in the diagonalization of the $Q-$Baxter operator \cite{bXXZ,MMR,CRS}.} were recovered within the representation theory of the $q-$Onsager algebra \cite{BK1,BK3}. Namely, in \cite{BK3} the spectral problem of the Hamiltonian was studied using certain properties of the $q-$Onsager algebra, especially those related with the concept of tridiagonal pairs \cite{Ter03}. Taking this point of view, it implies that eigenstates of the Hamiltonian can be expressed in terms of orthogonal symmetric functions generalizing the Askey-Wilson polynomials, a new class of special functions that are currently investigated in the mathematical literature. Let us also mention that some other properties exhibited in \cite{BK1} (see also \cite{BK}) - for instance the existence of $q-$deformed analogs of Davies's type of linear relations - may provide the starting point of another solution to the Hamiltonian's spectral problem by analogy with the solution at $q=1$ proposed in \cite{Davies}, a problem still unexplored. More generally, thanks to the recent progress in the classification of finite dimensional representations of the $q-$Onsager algebra and related algebraic structures (see \cite{IT} and references therein), a better understanding of the XXZ open spin chain or higher spins generalizations is clearly expected. However, besides the observation that there is still much room to be explored concerning finite size spin chains, the application of the Onsager's approach in the thermodynamic limit of lattice models - for instance the XXZ half-infinite spin chain - remained to be investigated. Several arguments motivate to consider this problem further:\vspace{1mm}

First, the transition from the finite size case to the thermodynamic limit in lattice models has been considered either in the context of the vertex operator approach or in the algebraic Bethe ansatz approach. Such analysis has not been carried out yet within the Onsager's approach, a problem that is closely related with the explicit construction of infinite dimensional representations of the $q-$Onsager and augmented $q-$Onsager algebras\footnote{Both algebras can be seen as special cases of the tridiagonal and augmented tridiagonal algebras which definitions can be found in \cite{Ter03} and \cite{IT}, respectively.}, as we are going to see. In the present article, for the first time  it is shown that the half-infinite XXZ spin chain for {\it any} type of boundary conditions (diagonal, non-diagonal, special cases) can be formulated using the $O_q(\widehat{sl_2})$ current algebra discovered in \cite{BSh1}. According to the choice of boundary conditions, the first modes of the currents are related with the generators of the $q-$Onsager and augmented $q-$Onsager algebras and act on infinite dimensional representations that are described in details for $-1<q<0$ in Section 4.\vspace{1mm} 

Secondly, recall that much is known for the special case of the half-infinite XXZ spin chain with  {\it diagonal} boundary conditions: the spectral problem and the calculation of correlation functions have been studied in details either within the vertex operator approach \cite{JKKKM} or within the algebraic Bethe ansatz approach \cite{KKMNST}. For the case of {\it non-diagonal} boundary conditions, the situation has remained essentially problematic. Indeed, integral representations for correlation functions - even in the simplest cases - have been, up to now, out of reach: either $q-$boson realizations of vacuum states are not known explicitly, or solving the inverse problem within the Bethe ansatz remains complicated. In the present article, the alternative path followed applies to any type (diagonal, non-diagonal or special) of boundary conditions: the properties of the new current algebra $O_q(\widehat{sl_2})$  and its modes are used to derive explicit expressions for the vacuum and excited states in terms of $q-$bosons in the massive regime $-1<q<0$ of the spin chain. \vspace{1mm} 

Third,  recall that in the case of the XXZ spin chain with periodic boundary conditions, in the thermodynamic limit the $U_q(\widehat{sl_2})$ algebra emerges as a hidden non-Abelian symmetry of the Hamiltonian \cite{FM,Jim0,vertex}. For many years, identifying the hidden symmetry of the open XXZ spin chain for diagonal or non-diagonal integrable boundary conditions has remained an open problem. In this article, it is shown that within Onsager's framework the answer to this problem is actually straightforward. Indeed,  in Section 2 two different types of spectrum generating algebras, denoted ${\cal A}_q$ and ${\cal A}^{diag}_q$, will be used to formulate the transfer matrix of the {\it finite} XXZ open spin chain for any type of boundary conditions. Whereas the algebra ${\cal A}_q$ is known to be associated with the $q-$Onsager algebra \cite{BK1}, the new algebra ${\cal A}^{diag}_q$ (obtained by fixing some parameters to zero in ${\cal A}_q$) is found to be associated with the so-called augmented $q-$Onsager algebra recently introduced in \cite{IT}. In the last Section, for non-diagonal or diagonal boundary conditions, it will be observed that the spectrum generating algebra (${\cal A}_q$ or ${\cal A}^{diag}_q$, respectively) associated with the spin chain of {\it finite size} becomes the hidden non-Abelian symmetry of the Hamiltonian in the thermodynamic limit. The long standing question of the hidden non-Abelian symmetry of the half-infinite XXZ spin chain is then solved for any type of boundary conditions.\vspace{1mm}

Let us also make a few comments on some mathematical objects that are involved in order to build an Onsager's approach in the thermodynamic limit. For instance, an exact solution based on an Onsager's formulation of the half-infinite XXZ open spin chain for any type of boundary conditions requires: (i) to identify the current algebra associated with either non-diagonal or diagonal boundary conditions; (ii)  to construct explicit infinite dimensional representations that will provide a bosonization scheme for the currents and local operators. Let us first make some comments about (i). As we will see in Section 2, in the finite size case according to the choice of boundary conditions two different types of spectrum generating algebras denoted ${\cal A}_q$ and ${\cal A}^{diag}_q$ have to be considered. Although the current algebra  introduced in \cite{BSh1} - there denoted $O_q(\widehat{sl_2})$  - generates ${\cal A}_q$ and applies to the case of generic non-diagonal boundary conditions, up to minor changes the defining relations of the second current algebra associated with ${\cal A}^{diag}_q$ are actually strictly identical (see Section 3).  Indeed, it will be shown that the algebra ${\cal A}^{diag}_q$ is nothing but a specialization of the algebra ${\cal A}_q$. The only difference between the two current algebras being in a choice of homomorphism given in Section 3, for simplicity the two current algebras will be denoted $O_q(\widehat{sl_2})$  in both cases. About (ii), recall that a realization of the $O_q(\widehat{sl_2})$ currents in terms of operators satisfying a Faddeev-Zamolodchikov algebra was already exhibited in \cite{BB}. This suggests that  infinite dimensional representations ($q-$vertex operators) of the $O_q(\widehat{sl_2})$ current algebra should be related with  $U_q(\widehat{sl_2})$ $q-$vertex operators. In the present article, this issue is clarified in Section 4. In particular, using the coideal structure the $U_q(\widehat{sl_2})$ $q-$vertex operators  are shown to be intertwiners of the $q-$Onsager and augmented $q-$Onsager algebra\footnote{To our knowledge, these results provide the first non-trivial examples of infinite dimensional representations for these two algebras.}. As a by product, an infinite dimensional analog of the two eigenbasis exhibited in \cite{Ter03} - namely, states that diagonalize the fundamental operators of the $q-$Onsager algebra - is identified in Section 5.\vspace{1mm} 

Having in mind above mentioned comments, the purpose of this article is to study in details the half-infinite XXZ spin chain within the Onsager's approach for {\it any} type of boundary conditions. Here, we will mainly focus on the formulation of the model in terms of the current algebra $O_q(\widehat{sl_2})$ introduced in \cite{BSh1}, its explicit relation with ${\cal A}_q$ or ${\cal A}^{diag}_q$, and its straightforward application to the diagonalization - i.e. the derivation of the spectrum and eigenstates - of the Hamiltonian of the model. For diagonal boundary conditions, known results \cite{JKKKM} are recovered. For upper (or lower) non-diagonal boundary conditions, new results are obtained, giving an access to integral representations of correlation functions that will be considered separately. For generic boundary conditions, the similar analysis that will be discussed elsewhere  is briefly sketched in the last Section.\vspace{1mm}

This paper is organized as follows: In Section 2, the Onsager's presentation of the finite size XXZ open spin chain for {\it any} type of boundary conditions\footnote{Up to now, only the case of {\it generic} non-diagonal boundary conditions has been considered within an Onsager's approach. See for instance \cite{BK,BK1}.} is considered in details. For generic non-diagonal boundary conditions, it is first reminded (see \cite{BK1} for details) how the transfer matrix can be explicitly written in terms of the elements of a spectrum generating algebra denoted ${\cal A}_q$. If some of the non-diagonal boundary parameters are set to zero, it is shown that the formulation remains essentially similar. However, the corresponding spectrum generating algebra is different, and denoted ${\cal A}^{diag}_q$. Using it, the description of the case of diagonal boundary conditions is proposed, which completes the formulation of \cite{BK3}. Explicit expressions of the elements of the spectrum generating algebras are reported in Appendix A. The relation between ${\cal A}_q$ and ${\cal A}^{diag}_q$ and algebras that appeared in the recent mathematical literature \cite{Ter03,IT} is then considered: It is already known that the first modes of ${\cal A}_q$  generate the $q-$Onsager algebra \cite{BK}. Here, we complete the analysis by showing that the first modes of ${\cal A}^{diag}_q$ generate the so-called augmented $q-$Onsager algebra recently introduced in \cite{IT}. In Section 3, the thermodynamic limit of the model is considered for {\it any} type of boundary conditions: sending one of the boundary to infinity, it is shown that the transfer matrix of the half-infinite XXZ open spin chain can be simply expressed in terms of $O_q(\widehat{sl_2})$ currents for $q$ generic. Whereas the exact relation between the $O_q(\widehat{sl_2})$ current algebra and the spectrum generating algebra ${\cal A}_q$ was given in \cite{BSh1}, here the homomorphism that relates $O_q(\widehat{sl_2})$ and ${\cal A}^{diag}_q$ is presented. By analogy with Section 2, the properties of the first modes are then considered in details in relation with two different coideal subalgebras of $U_q(\widehat{sl_2})$. Based on these, for $-1<q<0$ level one infinite dimensional representations of $O_q(\widehat{sl_2})$ are constructed in Section 4, where explicit expressions for the currents in terms of $U_q(\widehat{sl_2})$ $q-$vertex operators independently confirm the proposal of \cite{BB}. In Section 5, by analogy with the strategy applied in \cite{BK3} the spectral problem for two of the $O_q(\widehat{sl_2})$ currents is considered. Using this result, the diagonalization of the transfer matrix is then studied: for the case of diagonal boundary conditions, known results of Jimbo {\it et al.} \cite{JKKKM} are recovered and interpreted in light of the representation theory of the $O_q(\widehat{sl_2})$ current algebra. Then, the case of upper (or lower) non-diagonal boundary conditions is solved for the first time: whereas the spectrum is identical to the one for the diagonal case, an explicit expression for the eigenstates as an infinite sum is obtained. Few comments are added in the last Section. For instance, it is shown that the $q-$Onsager algebra or augmented $q-$Onsager algebra emerge as the non-Abelian symmetry of the thermodynamic limit of the XXZ open spin chain, according to the choice of boundary conditions. Although the existence of an infinite dimensional non-Abelian symmetry was definitely expected in the thermodynamic limit, we could not find any reference where it would be exhibited. Here this issue is definitely clarified. Finally, we point out some interesting phenomena for the special boundary conditions chosen in \cite{PS}.\vspace{1mm} 

In Appendix A, realizations of ${\cal A}_q$ and ${\cal A}^{diag}_q$  are given. In Appendix B, the Drinfeld-Jimbo presentation of $U_q(\widehat{sl_2})$ as well as basic definitions and objects that are used in the present article are recalled. In Appendix C, properties of $U_q(\widehat{sl_2})$ $q-$vertex operators and $q-$boson realizations that are used in Section 4 and 5 are given.
\begin{notation}
In this paper, we introduce the $q-$commutator $\big[X,Y\big]_q=qXY-q^{-1}YX$ where $q$ is the deformation parameter, assumed not to be a root of unity. 
\end{notation}

\section{Alternative presentations of the finite XXZ open spin chain}
In this Section, we first recall the Onsager's approach formulation of the XXZ open spin chain with generic boundary conditions \cite{BK1} (see also \cite{BK}). Then, we extend the formulation to the case of diagonal or special boundary conditions, preparing all necessary ingredients for studying the thermodynamic limit for any type of boundary conditions in further Sections.\vspace{1mm}

The finite size XXZ open spin chain with general integrable boundary conditions is the subject of numerous investigations in recent years. Starting from Sklyanin's work \cite{Skly88} for the special case of diagonal boundary conditions, it has been later on studied for {\it generic} or {\it special} (left-right related) boundary conditions and $q$ (root of unity) \cite{bXXZ,CRS,Galleas,Nic}. For general integrable boundary conditions and $q$, its Hamiltonian is given by: 
\beqa
H^{(N)}_{XXZ}&=&\sum_{k=1}^{N-1}\Big(\sigma_1^{k+1}\sigma_1^{k}+\sigma_2^{k+1}\sigma_2^{k} + \Delta\sigma_3^{k+1}\sigma_3^{k}\Big) +\ \frac{(q-q^{-1})}{2}\frac{(\epsilon_+ - \epsilon_-)}{(\epsilon_+ + \epsilon_-)}\sigma^1_3 + \frac{2}{(\epsilon_+ + \epsilon_-)}\big(k_+\sigma^1_+ + k_-\sigma^1_-\big)     \label{HN}\\
 &&\qquad\qquad  \qquad\qquad \qquad\qquad \qquad\qquad \ +\ \frac{(q-q^{-1})}{2}\frac{(\bepsilon_+ - \bepsilon_-)}{(\bepsilon_+ +\bepsilon_-)}\sigma^N_3 + \frac{2}{(\bepsilon_+ + \bepsilon_-)}\big(\bak_+\sigma^N_+ + \bak_-\sigma^N_-\big)    \ ,\nonumber
\eeqa
where $\sigma_{1,2,3}$ and $\sigma_\pm=(\sigma_1\pm i\sigma_2)/2$ are usual Pauli matrices. Here, $\Delta=(q+q^{-1})/2$\ \ denotes the anisotropy parameter and $\epsilon_\pm,k_\pm$ (resp. $ \bepsilon_\pm,\bak_\pm$) denote that the right (resp. left) boundary parameters associated with the right (resp. left) boundary. Considering a gauge transformation, note that one parameter might be removed. For symmetry reasons, we however keep the boundary parametrization as defined above. Restricting the parameters to special values or certain relations, one obtains the cases considered in \cite{ABBBQ,Skly88,PS,bXXZ,Rittenberg,Doikou,KKMNST,CRS}. \vspace{1mm}
   
In the literature, the most standard presentation of the XXZ open spin chain is based on a generalization of the quantum inverse problem to integrable systems with boundaries. In this approach, starting from an $R-$matrix acting on a finite dimensional representation, Hamiltonians of quantum integrable models are basically generated from solutions $K_-(\zeta)$, $K_+(\zeta)$ of the reflection and dual reflection equations, respectively \cite{Skly88}. In this standard presentation, the transfer matrix associated with the XXZ open spin chain  (\ref{HN}) can be written as:
\beqa
t^{(N)}(\zeta)=\frac{(-1)^N}{(\zeta^2+\zeta^{-2}-q^2-q^{-2})^N}\  tr_0\Big[K_+(\zeta){\bar R}_{0\verb"N"}(\zeta)\cdot\cdot\cdot
{\bar R}_{0\verb"1"}(\zeta)K_-(\zeta){\bar R}_{0\verb"1"}(\zeta)\cdot\cdot\cdot
{\bar R}_{0\verb"N"}(\zeta)  ]\Big]\ ,\label{tXXZ}
\eeqa
where
$tr_{0}$ denotes the trace over the two-dimensional auxiliary space,
\beqa {{\bar R}}(\zeta) =\left(
\begin{array}{cccc}
 \zeta q- \zeta^{-1}q^{-{1}}    & 0 & 0 & 0 \\
 0 & \zeta- \zeta^{-1} & q- q^{-{1}} & 0\\
0 & q-q^{-1} &  \zeta- \zeta^{-1} & 0  \\
0 & 0 & 0 & \zeta q- \zeta^{-1}q^{-{1}} 
\end{array} \right) \ ,\label{R}
\eeqa
and the most general elements\footnote{Note that $K_+(\zeta)=-K^t_-(-\zeta^{-1} q^{-1})|_{\epsilon_\pm\rightarrow \bepsilon_\mp; k_\pm\rightarrow \bak_\mp}$. } $K_\pm(\zeta)$ with $c-$number entries take the form
\beqa K_-(\zeta) &=&\left(
\begin{array}{cc} 
 \zeta \epsilon_{+} +  \zeta^{-1}\epsilon_{-}     &   k_+(\zeta^2-\zeta^{-2})/(q-q^{-1})\\
k_-(\zeta^2-\zeta^{-2})/(q-q^{-1})   &  \zeta \epsilon_{-} +  \zeta^{-1}\epsilon_{+}  \\
\end{array} \right) \ ,\label{K-}\\
K_+(\zeta) &=&\left(
\begin{array}{cc} 
 q\zeta\bepsilon_{+} +  q^{-1}\zeta^{-1}\bepsilon_{-}     &   \bak_+(q^2\zeta^2-q^{-2}\zeta^{-2})/(q-q^{-1})\\
\bak_-(q^2\zeta^2-q^{-2}\zeta^{-2})/(q-q^{-1})  &  q\zeta \bepsilon_{-} +  q^{-1}\zeta^{-1}\bepsilon_{+}  \\
\end{array} \right) \ \label{K+}\ .
\eeqa
In this formulation, the Hamiltonian of the XXZ open spin chain with general integrable boundary conditions (\ref{HN}) is obtained as follows\footnote{The identity operator acting on $N$ sites is denoted $I\!\!I^{(N)}=I\!\!I\otimes \cdot \cdot \cdot \otimes I\!\!I$.}:
\beqa
\frac{d}{d\zeta} ln(t^{(N)}(\zeta))|_{\zeta=1}=  \frac{2}{(q-q^{-1})}H^{(N)}_{XXZ} + \Big(\frac{(q-q^{-1})}{(q+q^{-1})} + \frac{2N}{(q-q^{-1})}\Delta \Big)I\!\!I^{(N)} \label{expH}\ .
\eeqa
More generally, higher mutually commuting local conserved quantities, say $H_n$ with $H_1\equiv H^{(N)}_{XXZ}$, can be derived similarly by taking higher derivatives of the transfer matrix (\ref{tfin}).
\vspace{2mm}

An alternative presentation of the XXZ open spin chain has been proposed in \cite{BK1}. It is inspired by the strategy developed by Onsager for the two-dimensional Ising model \cite{Ons44}, later works on the superintegrable chiral Potts and XY models \cite{Potts,PeAMT,Davies,Ar} (see also \cite{Ahn}) and the vertex operators approach \cite{vertex,JKKKM,Ko}: starting from the spectrum generating algebra or hidden non-Abelian algebra symmetry of a quantum integrable model, one is looking for the solution of the model solely using the representation theory of this algebra. In particular, such type of approach applies to the XXZ open spin chain which integrability condition can be associated with a $q-$deformed analog of the Onsager algebra for {\it generic} boundary conditions, as shown in \cite{BK1}. In this formulation, the transfer matrix can be written in terms of mutually commuting quantities ${{\cal I}}_{2k+1}^{(N)}$ that generate a $q-$deformed analog of the Onsager-Dolan-Grady's hierarchy\footnote{The Onsager's (also called Dolan-Grady) hierarchy is an Abelian algebra with elements of the form ${{I_{2n+1}}}= \bepsilon_+ (A_n+A_{-n}) + \bepsilon_-(A_{n+1}+A_{-n+1}) + \kappa G_{n+1}$ with $n\in{\mathbb Z}_+$ generated from the Onsager algebra with defining relations $\big[A_{n},A_{m}\big]= 4G_{n-m},\ \big[G_m,A_n\big] =2A_{n+m}-2A_{n-m}$ and $\ \big[G_n,G_m\big] =0$ for any $n,m\in{\mathbb Z}$\ .}. Namely,
\beqa
t_{gen-gen}^{(N)}(\zeta)= \sum_{k=0}^{N-1}{\cal F}_{2k+1}(\zeta)\ {{\cal I}}_{2k+1}^{(N)} + {\cal F}_0(\zeta)\ I\!\!I^{(N)}\quad \mbox{with} \qquad  \big[{\cal I}^{(N)}_{2k+1},{\cal I}^{(N)}_{2l+1}\big]=0 \qquad \label{sub} \label{tfin} 
\eeqa
for all $k,l\in 0,...,N-1$ where
\beqa
{{\cal I}}_{2k+1}^{(N)}=\bepsilon_+{\cal W}^{(N)}_{-k} + \bepsilon_- {\cal W}^{(N)}_{k+1} + \frac{1}{q^2-q^{-2}}\Big(\frac{\bak_-}{k_-} {\cal G}^{(N)}_{k+1} 
+ \frac{\bak_+}{k_+}{\tilde{\cal G}}^{(N)}_{k+1}\Big)\ \label{Ifin}
\eeqa
and ${\cal F}_{2k+1}(\zeta)$ are Laurent polynomials in $U(\zeta)=(q\zeta^2+q^{-1}\zeta^{-2})/(q+q^{-1})$. We refer the reader to \cite{BK1} for details. Note that the parameters $\epsilon_\pm$ of the right boundary - which do not appear explicitly in above formula - are actually hidden in the definition of the elements  ${\cal W}^{(N)}_{-k},{\cal W}^{(N)}_{k+1},{\cal G}^{(N)}_{k+1},{\tilde{\cal G}}^{(N)}_{k+1}$ of the spectrum generating algebra which explicit expressions are recalled in Appendix A.  

\begin{rem} Another Onsager's presentation can be alternatively considered , in which case the elements of the spectrum generating algebra contain the parameters $\bepsilon_\pm,\bak_\pm$ of the left boundary. In this case, one substitutes in (\ref{tfin}) 
\beqa
{{\cal I}}_{2k+1}^{(N)}\rightarrow{\overline{\cal I}}_{2k+1}^{(N)} \quad \mbox{where}\quad {\overline{\cal I}}_{2k+1}^{(N)}= \epsilon_+ \overline{\cal W}^{(N)}_{k+1} + \epsilon_-\overline{\cal W}^{(N)}_{-k} + \frac{1}{q^2-q^{-2}}\Big(\frac{k_-}{\bak_-} \overline{\tilde{\cal G}}^{(N)}_{k+1} + \frac{k_+}{\bak_+} \overline{\cal G}^{(N)}_{k+1}\Big)\ .\label{Ifindual}
\eeqa
The functions $\overline{\cal F}_{2k+1}(\zeta),\overline{\cal F}_{0}(\zeta)$ can be derived following \cite{BK1}. Here, the elements are given by:
\beqa
\label{elB1}\overline{\cal W}^{(N)}_{-k} = \Pi_N\big({\cal W}^{(N)}_{k+1}\big)|_{\epsilon_\pm\rightarrow\bepsilon_\pm;k_\pm\rightarrow\ \bak_\pm}\ ,\quad  \overline{\cal W}^{(N)}_{k+1} = \Pi_N\big({\cal W}^{(N)}_{-k}\big)|_{\epsilon_\pm\rightarrow\bepsilon_\pm;k_\pm\rightarrow\ \bak_\pm}\ , \\
\overline{\cal G}^{(N)}_{k+1} = \Pi_N\big(\tilde{\cal G}^{(N)}_{k+1}\big)|_{\epsilon_\pm\rightarrow\bepsilon_\pm;k_\pm\rightarrow\ \bak_\pm}\ ,\quad  \overline{\tilde{\cal G}}^{(N)}_{k+1} = \Pi_N\big({\cal G}^{(N)}_{k+1}\big)|_{\epsilon_\pm\rightarrow\bepsilon_\pm;k_\pm\rightarrow\ \bak_\pm}\ ,\nonumber
\eeqa
where the permutation operator $\Pi_N\big(a_1\otimes a_2\otimes ... \otimes a_N\big) = a_N\otimes ... \otimes a_2 \otimes a_1$ is used.
\end{rem}\vspace{2mm} 

Having recalled the Onsager's type of presentation (\ref{tfin}) for the XXZ open spin chain with {\it generic} boundary conditions,
a natural question is whether such presentation also exists for special boundary conditions that have been considered in the literature.  Actually, for special {\it right diagonal} boundary conditions it is also the case provided certain changes in the definition of the basic objects: Following the analysis of \cite{BK1}, it is easy to show that the {\it diagonal} boundary analog of the $q-$Dolan-Grady hierarchy is associated with an Abelian subalgebra generated by the elements ${\cal J}^{(N)}_{2l+1}$ such that 
\beqa
{{\cal J}}_{2k+1}^{(N)}=\bepsilon_+{\cal K}^{(N)}_{-k} + \bepsilon_- {\cal K}^{(N)}_{k+1} + \frac{1}{q^2-q^{-2}}\Big(  \bak_-{\cal Z}^{(N)}_{k+1} 
+ \bak_+{\tilde{\cal Z}}^{(N)}_{k+1} \Big)\ \label{Jfin}
\eeqa
where the explicit expressions of ${\cal K}^{(N)}_{-k},{\cal K}^{(N)}_{k+1},{\cal Z}^{(N)}_{k+1},{\tilde{\cal Z}}^{(N)}_{k+1}$ are reported in Appendix A. In this special case, the transfer matrix associated with the Hamiltonian (\ref{HN}) for $k_\pm\equiv 0$ takes the form:
\beqa
t_{gen-diag}^{(N)}(\zeta)= \sum_{k=0}^{N-1}{\cal F}^{diag}_{2k+1}(\zeta)\ {{\cal J}}_{2k+1}^{(N)} + {\cal F}^{diag}_0(\zeta)\ I\!\!I^{(N)}\quad \mbox{with} \qquad  \big[{\cal J}^{(N)}_{2k+1},{\cal J}^{(N)}_{2l+1}\big]=0 \ .\label{sub} \label{tfingd} 
\eeqa

\begin{rem} For special {\it left diagonal} (but right generic) boundary conditions, the transfer matrix $t_{diag-gen}^{(N)}(\zeta)$ can be aternatively presented in terms of
\beqa
{\overline{\cal J}}_{2k+1}^{(N)}=\epsilon_+ \overline{\cal K}^{(N)}_{k+1} + \epsilon_-\overline{\cal K}^{(N)}_{-k} +  \frac{1}{q^2-q^{-2}}\Big(k_- \overline{\tilde{\cal Z}}^{(N)}_{k+1} + k_+ \overline{\cal Z}^{(N)}_{k+1} \Big)\ \label{Ifindualaug}
\eeqa
where 
\beqa
\label{elB2} \overline{\cal K}^{(N)}_{-k} = \Pi_N\big({\cal K}^{(N)}_{k+1}\big)|_{\epsilon_\pm\rightarrow\bepsilon_\pm}\ ,\quad  \overline{\cal K}^{(N)}_{k+1} = \Pi_N\big({\cal K}^{(N)}_{-k}\big)|_{\epsilon_\pm\rightarrow\bepsilon_\pm}\ ,\\
\overline{\cal Z}^{(N)}_{k+1} = \Pi_N\big(\tilde{\cal Z}^{(N)}_{k+1}\big)|_{\epsilon_\pm\rightarrow\bepsilon_\pm}\ ,\quad  \overline{\tilde{\cal Z}}^{(N)}_{k+1} = \Pi_N\big({\cal Z}^{(N)}_{k+1}\big)|_{\epsilon_\pm\rightarrow\bepsilon_\pm}\ \nonumber.
\eeqa
\end{rem} 

As a consequence, the transfer matrix $t_{diag-diag}^{(N)}(\zeta)$ for the special case of {\it left and right diagonal} boundary conditions - the simplest case studied in \cite{Skly88} using the presentation (\ref{tXXZ}) - also admits an Onsager's type of presentation, where mutually commuting quantities are simply given by  (\ref{Jfin}) with $\bak_\pm\equiv 0$ or, alternatively, (\ref{Ifindualaug}) with $k_\pm\equiv 0$.
\vspace{2mm}

For each type of boundary conditions, let us now describe some basic aspects of the corresponding spectrum generating algebras. According to the choice of boundary conditions - generic non-diagonal or generic diagonal - associated with the left and right side of the spin chain, two different types of spectrum generating algebras arise in the Onsager's presentation of the XXZ open spin chain. 
\vspace{1mm}

$\bullet$ {\bf Parameters $k_\pm\neq 0$}: The elements ${\cal W}^{(N)}_{-k},{\cal W}^{(N)}_{k+1},{\cal G}^{(N)}_{k+1},{\tilde{\cal G}}^{(N)}_{k+1}$  in (\ref{Ifin}) are known to satisfy the defining relations of the infinite dimensional $q-$deformed analog of the Onsager algebra ${\cal A}_q$ introduced in \cite{BK} (see \cite{BK1}  for details) which ensures the integrability of the model (\ref{HN}). Recall that the defining relations of  ${\cal A}_q$ are given by:
\beqa
&&\label{qOns} \big[{\textsf W}_0,{\textsf W}_{k+1}\big]=\big[{\textsf W}_{-k},{\textsf W}_{1}\big]=\frac{1}{(q^{1/2}+q^{-1/2})}\big({\tilde{\textsf G}_{k+1} } - {{\textsf G}_{k+1}}\big)\ ,\\
&&\big[{\textsf W}_0,{\textsf G}_{k+1}\big]_q=\big[{\tilde{\textsf G}}_{k+1},{\textsf W}_{0}\big]_q=\rho{\textsf W}_{-k-1}-\rho{\textsf W}_{k+1}\ ,\nonumber\\
&&\big[{\textsf G}_{k+1},{\textsf W}_{1}\big]_q=\big[{\textsf W}_{1},{\tilde{\textsf G}}_{k+1}\big]_q=\rho{\textsf W}_{k+2}-\rho{\textsf W}_{-k}\ ,\nonumber\\
&&\big[{\textsf W}_0,{\textsf W}_{-k}\big]=0\ ,\quad 
\big[{\textsf W}_1,{\textsf W}_{k+1}\big]=0\ ,\quad \nonumber\\
&&\big[{\textsf G}_{k+1},{\textsf G}_{l+1}\big]=0\ ,\quad   \big[{\tilde{\textsf G}}_{k+1},\tilde{{\textsf G}}_{l+1}\big]=0\ ,\quad
\big[{\tilde{\textsf G}}_{k+1},{\textsf G}_{l+1}\big]
+\big[{{\textsf G}}_{k+1},\tilde{{\textsf G}}_{l+1}\big]=0\ ,\nonumber
\eeqa
with $k,l\in {\mathbb N}$. For the finite dimensional tensor product representation associated with (\ref{HN}), the elements act as ${\cal W}^{(N)}_{-k},{\cal W}^{(N)}_{k+1},{\cal G}^{(N)}_{k+1},{\tilde{\cal G}}^{(N)}_{k+1}$ given in Appendix A and we have the sustitution \cite{BK1}
\beqa
\rho =(q+q^{-1})^2 k_+k_-\ .\label{rho}
\eeqa

By analogy with the situation for integrable models associated with the Onsager algebra (i.e. the undeformed case), for instance the Ising and superintegrable Potts models, considering a finite dimensional space on which the Hamiltonian (\ref{HN}) acts implies the existence of additionnal relations among the generators. Such relations which are $q-$deformed analogs of Davis' relations \cite{Davies} have been derived in \cite{BK1} (see also \cite{BK})  for the model (\ref{HN}). Explicitly, they take the form:
\beqa
\label{c4}&&-\frac{(q-q^{-1})}{k_+k_-}\omega_0^{(N)}\cW_{0}^{(N)}+\sum^{N}_{k=1}C_{-k+1}^{(N)}\cW_{-k}^{(N)} + \epsilon^{(N)}_{+}I\!\!I^{(N)}=0\ ,\\
&&-\frac{(q-q^{-1})}{k_+k_-}\omega_0^{(N)}\cW_{1}^{(N)}+\sum^{N}_{k=1}C_{-k+1}^{(N)}\cW_{k+1}^{(N)} + \epsilon^{(N)}_{-}I\!\!I^{(N)}=0\ ,\nonumber \\
&&-\frac{(q-q^{-1})}{k_+k_-}\omega_0^{(N)}\cG_{1}^{(N)}+\sum^{N}_{k=1}C_{-k+1}^{(N)}\cG_{k+1}^{(N)}=0\ ,\nonumber\\
&&-\frac{(q-q^{-1})}{k_+k_-}\omega_0^{(N)}{\tilde \cG}_{1}^{(N)}+\sum^{N}_{k=1}C_{-k+1}^{(N)}{\tilde \cG}_{k+1}^{(N)}=0\ ,\nonumber\
\eeqa
where the explicit expressions for the coefficients $\omega_0^{(N)},C_{-k+1}^{(N)},\epsilon^{(N)}_\pm$ in terms of $k_\pm,\epsilon_\pm,q,N$  are given in \cite{BK1}. Strictly speaking, for generic parameters  the spectrum generating algebra associated with (\ref{HN}) is the quotient of the infinite dimensional algebra ${\cal A}_q$ by the set of relations (\ref{c4}).\vspace{1mm}

For non-vanishing parameters $k_\pm$, let us also make some important comments about the relation between the algebra  ${\cal A}_q$ and the so-called $q-$Onsager algebra exhibited in \cite{BK}, which will be useful in the analysis of further Sections. Remarkably, according to the explicit expressions  (\ref{rep0})  for the first elements and {\it generic} values of $\epsilon_\pm,k_\pm,q$, 
\beqa
{\cal W}^{(N)}_0&=& (k_+\sigma_+ + k_-\sigma_-)\otimes I\!\!I^{(N-1)} + q^{\sigma_3}\otimes {\cal W}_0^{(N-1)}\ ,\quad {\cal W}^{(0)}_0=\epsilon_+ \ ,\label{op} \\
{\cal W}^{(N)}_1&=& (k_+\sigma_+ + k_-\sigma_-)\otimes I\!\!I^{(N-1)} + q^{-\sigma_3}\otimes {\cal W}_1^{(N-1)}\ ,\quad {\cal W}^{(0)}_1= \epsilon_-\ ,\nonumber
\eeqa
one observes\footnote{More generally, all higher elements can be written as polynomials in ${\cal W}^{(N)}_0,{\cal W}^{(N)}_1$ \cite{BB3}. For instance, ${\cal G}^{(N)}_1= I\!\!I\otimes {\cal G}^{(N-1)}_1 + (q^2-q^{-2})k_-\sigma_-\otimes \left({\cal W}_0^{(N-1)}+{\cal W}_1^{(N-1)}\right) + k_+k_-(q-q^{-1})I\!\!I^{(N)}=[{\cal W}^{(N)}_1,{\cal W}^{(N)}_0]_q$. Similarly, one has ${\tilde{\cal G}}^{(N)}_1=[{\cal W}^{(N)}_0,{\cal W}^{(N)}_1]_q$. This gives an explicit relation between the generators of ${\cal A}_q$ in terms of the ones of the $q-$Onsager algebra for the finite dimensional representations here considered.} that they satisfy
the defining relations of the $q-$Onsager algebra, the so-called $q-$Dolan-Grady relations \cite{Ter03}:
\beqa \big[{\cal W}^{(N)}_0,\big[{\cal W}^{(N)}_0,\big[{\cal W}^{(N)}_0,{\cal
W}^{(N)}_1\big]_q\big]_{q^{-1}}\big]&=&\rho\big[{\cal W}^{(N)}_0,{\cal W}^{(N)}_1\big]\ ,\label{Talg}\\
\big[{\cal W}^{(N)}_1,\big[{\cal W}^{(N)}_1,\big[{\cal W}^{(N)}_1,{\cal
W}^{(N)}_0\big]_q\big]_{q^{-1}}\big]&=&\rho\big[{\cal W}^{(N)}_1,{\cal W}^{(N)}_0\big]\ .\nonumber
\eeqa
Note that these relations are a special case of the defining relations of the tridiagonal algebras \cite{Ter03}. In the next Section, we will argue that the spectrum generating algebra associated with the half-infinite XXZ spin chain for generic {\it non-diagonal} boundary conditions is ${\cal A}_q$, which first elements satisfy (\ref{Talg}). Note that for infinite dimensional representations which are relevant in the thermodynamic limit of (\ref{HN}), the additional relations (\ref{c4}) do not arise \cite{BSh1}. \vspace{2mm}

$\bullet$ {\bf Parameters $k_\pm\equiv 0$}: We now turn to the spectrum generating algebra which is relevant for the study of  the Hamiltonian (\ref{HN}) with generic {\it diagonal} boundary conditions, i.e. for $k_\pm=0$, which implies $\rho=0$. By analogy with the analysis above, the defining relations of the infinite dimensional algebra ${\cal A}^{diag}_q$ satisfied by the elements ${\cal K}^{(N)}_{-k}$, ${\cal K}^{(N)}_{k+1}$, ${\cal Z}^{(N)}_{k+1}$, ${\tilde{\cal Z}}^{(N)}_{k+1}$ which ensures the integrability of the model (\ref{HN}) for {\it right diagonal} boundary conditions, as well as the linear relations similar to (\ref{c4}), can be derived using the substitutions (\ref{subsKZ}) in (\ref{qOns}), (\ref{c4}) and setting $k_\pm=0$. Similarly to the case of generic boundary conditions, for our purpose it will be however sufficient to focus on the set of relations satisfied by the first elements. Using the explicit expressions  (\ref{rep}), one has:
\beqa
{\cal K}^{(N)}_0&=& q^{\sigma_3}\otimes {\cal K}_0^{(N-1)}\ ,\quad {\cal K}^{(0)}_0=\epsilon_+\ ,\label{opdiag} \\
{\cal K}^{(N)}_1&=& q^{-\sigma_3}\otimes {\cal K}_1^{(N-1)}\ ,\quad {\cal K}^{(0)}_1=\epsilon_-\ ,\nonumber \\
{\cal Z}_{1}^{(N)}&=& I\!\!I \otimes {\cal Z}_{1}^{(N-1)} + (q^2-q^{-2})\sigma_-\otimes\left({\cal K}^{(N-1)}_0+{\cal K}^{(N-1)}_1\right)\ ,\nonumber \\
\tilde{\cal Z}_{1}^{(N)}&=& I\!\!I \otimes \tilde{\cal Z}_{1}^{(N-1)} + (q^2-q^{-2})\sigma_+\otimes\left({\cal K}^{(N-1)}_0+{\cal K}^{(N-1)}_1\right)\ ,\quad {\cal Z}^{(0)}_1={\tilde{\cal Z}}^{(0)}_1=0\ .\nonumber
\eeqa
By straightforward calculations, for generic values of $\epsilon_\pm,q$ the elements  ${\cal K}^{(N)}_{0},{\cal K}^{(N)}_{1},{\cal Z}^{(N)}_{1},{\tilde{\cal Z}}^{(N)}_{1}$ are found to generate the {\it augmented $q-$Onsager algebra} with defining relations:
\beqa
\big[ {\cal K}^{(N)}_0,{\cal K}^{(N)}_1\big]&=&0\ ,\label{Taug}\\
{\cal K}^{(N)}_0{\cal Z}^{(N)}_1&=&q^{-2} {\cal Z}^{(N)}_1{\cal K}^{(N)}_0\ ,\qquad {\cal K}^{(N)}_0\tilde{\cal Z}^{(N)}_1=q^{2}\tilde{\cal Z}^{(N)}_1{\cal K}^{(N)}_0\ ,\nonumber\\
{\cal K}^{(N)}_1{\cal Z}^{(N)}_1&=&q^{2}{\cal Z}^{(N)}_1{\cal K}^{(N)}_1\ ,\qquad{\cal K}^{(N)}_1\tilde{\cal Z}^{(N)}_1=q^{-2}\tilde{\cal Z}^{(N)}_1{\cal K}^{(N)}_1\ ,\nonumber\\
\big[{\cal Z}^{(N)}_1,\big[{\cal Z}^{(N)}_1,\big[{\cal Z}^{(N)}_1,\tilde{\cal
Z}^{(N)}_1\big]_q\big]_{q^{-1}}\big]&=&\rho_{diag}{\cal Z}^{(N)}_1(\,{\cal K}^{(N)}_1{\cal K}^{(N)}_1-\,{\cal K}^{(N)}_0{\cal K}^{(N)}_0){\cal Z}^{(N)}_1,\nonumber\\
\big[\tilde{\cal Z}^{(N)}_1,\big[\tilde{\cal Z}^{(N)}_1,\big[\tilde{\cal Z}^{(N)}_1,{\cal
Z}^{(N)}_1\big]_q\big]_{q^{-1}}\big]&=&\rho_{diag}\tilde{\cal Z}^{(N)}_1({\cal K}^{(N)}_0{\cal K}^{(N)}_0-{\cal K}^{(N)}_1{\cal K}^{(N)}_1)\tilde{\cal Z}^{(N)}_1\  
\nonumber\eeqa
with 
\beqa
\rho_{diag}=\frac{(q^3-q^{-3})(q^2-q^{-2})^3}{q-q^{-1}}\  .\label{rhodiag}
\eeqa
Note that the augmented $q-$Onsager algebra is a special case of the augmented tridiagonal algebra \cite{IT}, which finite dimensional representations for $q$ not a root of unity have been classified in \cite{IT}.  In the next Section, we will argue that the spectrum generating algebra associated with the half-infinite XXZ spin chain for generic {\it diagonal} boundary conditions is ${\cal A}^{diag}_q$, which first elements satisfy (\ref{Taug}).\vspace{2mm}

In the next Section, the thermodynamic limit $N\rightarrow \infty$ of the Hamiltonian (\ref{HN}) will be considered in details. In this limit, the presentations of the transfer matrices $t^{(N)}(\zeta)$ of the form (\ref{tfin}) or (\ref{tfingd}) will be suitable for our purpose. These are linear combinations of  the mutually commuting quantities (\ref{Ifindual}) for {\it non-diagonal} boundary conditions and (\ref{Ifindualaug}) for {\it diagonal} boundary conditions. From the analysis above and the definitions (\ref{elB1}),  (\ref{elB2}), let us observe that the elements $\overline{\cal W}^{(N)}_{-k},\overline{\cal W}^{(N)}_{k+1},\overline{\cal G}^{(N)}_{k+1},\overline{\tilde{\cal G}}^{(N)}_{k+1}$ and $\overline{\cal K}^{(N)}_{-k},\overline{\cal K}^{(N)}_{k+1},\overline{\cal Z}^{(N)}_{k+1},\overline{\tilde{\cal Z}}^{(N)}_{k+1}$ also generate the infinite dimensional algebras ${\cal A}_q$ and ${\cal A}^{diag}_q$, respectively, and linear relations of the form (\ref{c4}).  In particular, the elements:
\beqa
\overline{\cal W}^{(N)}_0&=& I\!\!I^{(N-1)}\otimes (\bak_+\sigma_+ + \bak_-\sigma_-) + \overline{\cal W}_0^{(N-1)} \otimes q^{-\sigma_3}\ ,\quad \overline{\cal W}^{(0)}_0=\bepsilon_- \ ,\label{opbar} \\
\overline{\cal W}^{(N)}_1&=& I\!\!I^{(N-1)} \otimes (\bak_+\sigma_+ + \bak_-\sigma_-) + \overline{\cal W}_1^{(N-1)}\otimes q^{\sigma_3} \ ,\quad \overline{\cal W}^{(0)}_1= \bepsilon_+\ \nonumber
\eeqa
satisfy the $q-$Onsager algebra relations (\ref{Talg}) with $\rho =(q+q^{-1})^2 \bak_+\bak_-$.\vspace{1mm} 

On the other hand, the elements:
\beqa
\overline{\cal K}^{(N)}_0&=& \overline{\cal K}_0^{(N-1)} \otimes q^{-\sigma_3}\ ,\quad \overline{\cal K}^{(0)}_0=\bepsilon_-\ ,\label{opdiagbar}\\
\overline{\cal K}^{(N)}_1&=& \overline{\cal K}_1^{(N-1)} \otimes q^{\sigma_3}\ ,\quad \overline{\cal K}^{(0)}_1=\bepsilon_+\ ,\nonumber \\
\overline{\cal Z}_{1}^{(N)}&=& \overline{\cal Z}_{1}^{(N-1)} \otimes I\!\!I   + (q^2-q^{-2})\left(\overline{\cal K}^{(N-1)}_0 +\overline{\cal K}^{(N-1)}_1\right) \otimes \sigma_+ \ ,\nonumber \\
\overline{\tilde{\cal Z}}_{1}^{(N)}&=& \overline{\tilde{\cal Z}}_{1}^{(N-1)}  \otimes I\!\!I + (q^2-q^{-2})  \left(\overline{\cal K}^{(N-1)}_0 +\overline{\cal K}^{(N-1)}_1\right) \otimes \sigma_- \ ,\quad \overline{\cal Z}^{(0)}_1=\overline{\tilde{\cal Z}}^{(0)}_1=0\ \nonumber
\eeqa
satisfy the augmented $q-$Onsager algebra with relations (\ref{Taug}) and (\ref{rhodiag}). 

\vspace{2mm}

To resume, recall that the explicit relation between Sklyanin's presentation (\ref{tXXZ}) \cite{Skly88} and Onsager's type of presentation (\ref{tfin}) of the XXZ open spin chain with generic boundary conditions has been described in details  in \cite{BK1} (see also \cite{BK}). Above results for the special case of right diagonal, left diagonal or right and left diagonal boundary conditions complete the correspondence. According to the choice of boundary conditions, two different types of spectrum generating algebras have to be considered in this framework: either ${\cal A}_q$  or ${\cal A}^{diag}_q$, which first elements satisfy the $q-$Onsager algebra (\ref{Talg}) (see \cite{BK,BK1} for details) or the augmented $q-$Onsager algebra (\ref{Taug}) exhibited here, respectively. These results are collected in the following table, where the set of integrals of motions (IMs) are specified according to the presentation chosen: \vspace{2mm}

\centerline{
\begin{tabular}[t]{|c|c|c|c|} \hline 
       {\bf open XXZ chain} & Spectrum gen. algebra  
       	&  IMs (1st presentation)& IMs (2nd presentation)\\
	     \hline 
       right-left generic bcs. & ${\cal A}_q\rightarrow $ $q-$Onsager &  ${\cal I}^{(N)}_{2k+1}$ &  $\overline{\cal I}^{(N)}_{2k+1}$  \\
	     \hline
      right diag. bcs. $k_\pm= 0$      &  ${\cal A}^{diag}_q\rightarrow $ aug. $q-$Onsager &  ${\cal J}^{(N)}_{2k+1}$ &   \\
	     & {\it or} \ \ \  ${\cal A}_q\rightarrow $ $q-$Onsager & & $\overline{\cal I}^{(N)}_{2k+1}|_{k_\pm=0}$ \\
	     \hline
      left diag. bcs. $\bak_\pm= 0$      &  ${\cal A}_q\rightarrow $ $q-$Onsager & ${\cal I}^{(N)}_{2k+1}|_{\bak_\pm=0}$ & \\
	     &  {\it or} \ \ ${\cal A}^{diag}_q\rightarrow $ aug. $q-$Onsager  & & $\overline{\cal J}^{(N)}_{2k+1}$ \\
	     \hline
      right-left diag. bcs. $\bak_\pm=k_\pm= 0$      &  ${\cal A}^{diag}_q\rightarrow $ aug. $q-$Onsager &  ${\cal J}^{(N)}_{2k+1}|_{\bak_\pm=0}$ & $\overline{\cal J}^{(N)}_{2k+1}|_{k_\pm=0}$   \\
\hline 
\end{tabular}}
\vspace{6mm}

Note that the spectrum of the XXZ open spin chain Hamiltonian (\ref{HN}) with {\it right} diagonal boundary conditions ($k_\pm=0$) and {\it  left} diagonal boundary ($\bak_\pm=0$) conditions may be considered using the properties either of ${\cal A}_q$ or ${\cal A}^{diag}_q$. Also, it is important to stress that the list of cases presented above is not exhaustive: for instance, one may consider the set of diagonal boundary conditions $k_\pm=\bak_\pm=0$, $\epsilon_+\neq 0$, $\bepsilon_-\neq 0$ and $\epsilon_-=\bepsilon_+=0$ discussed in \cite{ABBBQ,PS}. In this special case, let us remark that the defining relations satisfied by the fundamental elements of the corresponding `larger' spectrum generating algebra can be derived in a straighforward manner (see a related work \cite{R}). As will be discussed in the last Section, in the thermodynamic limit the diagonalization of the $q-$Dolan-Grady hierarchy in this special case exhibits interesting features.\vspace{-1mm}

\section{Onsager's presentation: the thermodynamic limit}
The purpose of this Section is to show that, in the thermodynamic limit, the Onsager's formulation of the XXZ open spin chain for any type of integrable boundary conditions becomes rather simple: the transfer matrix can be written in terms of the elements of a current algebra denoted $O_q(\widehat{sl_2})$, which sligthly generalizes the current algebra introduced in \cite{BSh1}. According to the choice of parameters $k_\pm$, two different types of homomorphisms from $O_q(\widehat{sl_2})$ to ${\cal A}_q$ or ${\cal A}^{diag}_q$ are exhibited. Also, the relation with certain coideal subalgebras of $U_q(\widehat{sl_2})$ is established, that will play a central role in the next Section for the construction of level one infinite dimensional representations ($q-$vertex operators) of $O_q(\widehat{sl_2})$ starting from $U_q(\widehat{sl_2})$ ones.
\vspace{1mm} 

The half-infinite XXZ open spin chain with an integrable boundary can be considered as the thermodynamic limit $N\rightarrow \infty$ of the finite XXZ open spin chain (\ref{HN}). Consider the Hamiltonian:
\beqa
\qquad H_{\frac{1}{2}XXZ}&=&-\frac{1}{2}\sum_{k=1}^{\infty}\Big(\sigma_1^{k+1}\sigma_1^{k}+\sigma_2^{k+1}\sigma_2^{k} + \Delta\sigma_3^{k+1}\sigma_3^{k}\Big) - \frac{(q-q^{-1})}{4}\frac{(\epsilon_+-\epsilon_-)}{(\epsilon_++\epsilon_-)}\sigma^1_3 - \frac{1}{(\epsilon_++\epsilon_-)}\big(k_+\sigma^1_+ + k_-\sigma^1_-\big) 
\label{Hsemi}\  .
\eeqa
Note that the normalization in front of the Hamiltonian has been changed compared with (\ref{HN}), to fit later on with the definitions of \cite{JKKKM} for the special case of {\it diagonal} boundary conditions $k_\pm=0$. By definition, the Hamiltonian formally acts on an infinite dimensional vector space ${\cal V}$ which can be written as an infinite tensor product of $2-$dimensional  ${\mathbb C}^2$ vector space. According to the ordering of the tensor components in (\ref{Hsemi}),
\beqa
{\cal V}= \cdot \cdot\cdot \otimes {\mathbb C}^2 \otimes {\mathbb C}^2  \otimes {\mathbb C}^2 \ .\label{halfV}
\eeqa

A transfer matrix associated with the Hamiltonian (\ref{Hsemi}) can be proposed by analogy with the expressions (\ref{tfin}), (\ref{tfingd}) derived for the finite size case. As we are going to explain, it can be written in terms of the elements of the current algebra associated with ${\cal A}_q$ for $k_\pm\neq 0$ or ${\cal A}^{diag}_q$ for $k_\pm=0$.\vspace{1mm} 

First, recall that the defining relations of ${\cal A}_q$ can be derived from the current algebra that has been introduced in [\cite{BSh1}, Definition 2.2], well-defined for $k_\pm\neq 0$ . With minor changes, the defining relations of  ${\cal A}^{diag}_q$ with $k_\pm=0$ can be obtained similarly. Actually, both sets of defining relations follow from a slightly more general current algebra - denoted here $O_q(\widehat{sl_2})$ for simplicity - using two different homomorphisms (mode expansion) given below. To show this, following \cite{BSh1} define the formal variables $U(\zeta)=(q\zeta^2+q^{-1}\zeta^{-2})/(q+q^{-1})$. Let us introduce the current algebra $O_q(\widehat{sl_2})$ with defining relations:
\beqa
&&\big[{\cW}_\pm(\zeta),{\cW}_\pm(\xi)\big]=0\ ,\qquad\qquad\qquad\qquad\qquad\qquad\qquad\label{ec1}\\
&&\big[{\cW}_+(\zeta),{\cW}_-(\xi)\big]+\big[{\cW}_-(\zeta),{\cW}_+(\xi)\big]=0\ ,\qquad\qquad\qquad\qquad\qquad\qquad\qquad\label{ec3}\\
&&(U(\zeta)-U(\xi))\big[{\cW}_\pm(\zeta),{\cW}_\mp(\xi)\big]= \frac{(q-q^{-1})}{(q+q^{-1})^3}\left({\cal Z}_\pm(\zeta){\cal Z}_\mp(\xi)-{\cal Z}_\pm(\xi){\cal Z}_\mp(\zeta)\right)\ ,\qquad\qquad\qquad\label{ec4}\nonumber \\
&&{\cW}_\pm(\zeta){\cW}_\pm(\xi)-{\cW}_\mp(\zeta){\cW}_\mp(\xi)+\frac{1}{(q^2-q^{-2})(q+q^{-1})^2}\big[{\cal Z}_\pm(\zeta),{\cal Z}_\mp(\xi)\big]\qquad\qquad\qquad\qquad \qquad\label{ec5}\\
&&\qquad\qquad\qquad\qquad\qquad\qquad+ \ \frac{1-U(\zeta)U(\xi)}{U(\zeta)-U(\xi)}\big({\cW}_\pm(\zeta){\cW}_\mp(\xi)-{\cW}_\pm(\xi){\cW}_\mp(\zeta)\big)=0\ ,\nonumber\\
&&U(\zeta)\big[{\cal Z}_\mp(\xi),{\cW}_\pm(\zeta)\big]_q -U(\xi)\big[{\cal Z}_\mp(\zeta),{\cW}_\pm(\xi)\big]_q - (q-q^{-1})\big({\cW}_\mp(\zeta){\cal Z}_\mp(\xi)-{\cW}_\mp(\xi){\cal Z}_\mp(\zeta)\big) = 0\ ,\nonumber\\
&&U(\zeta)\big[{\cW}_\mp(\zeta),{\cal Z}_\mp(\xi)\big]_q -U(\xi)\big[{\cW}_\mp(\xi),{\cal Z}_\mp(\zeta)\big]_q - (q-q^{-1})\big({\cW}_\pm(\zeta){\cal Z}_\mp(\xi)-{\cW}_\pm(\xi){\cal Z}_\mp(\zeta)\big) = 0\label{ec7}\ ,\nonumber\\
&&\big[{\cal Z}_\epsilon(\zeta),{\cW}_\pm(\xi)\big]+\big[{\cW}_\pm(\zeta),{\cal Z}_\epsilon(\xi)\big]=0 \ ,\quad \forall \epsilon=\pm\label{ec8}\ ,\qquad\qquad\qquad\qquad\qquad\qquad\qquad\\
&&\big[{\cal Z}_\pm(\zeta),{\cal Z}_\pm(\xi)\big]=0\ ,\label{ec9}\qquad\qquad\qquad\qquad\qquad\qquad\qquad\\ 
&&\big[{\cal Z}_+(\zeta),{\cal Z}_-(\xi)\big]+\big[{\cal Z}_-(\zeta),{\cal Z}_+(\xi)\big]=0\ .\qquad\qquad\qquad\qquad\qquad\qquad\qquad\ \label{ec16}
\eeqa
The homomorphism proposed in [\cite{BSh1}, Theorem 2] gives the explicit relation between the $O_q(\widehat{sl_2})$ current algebra for $k_\pm\neq 0$ with defining relations  (\ref{ec1})-(\ref{ec16})  and the defining relations of  ${\cal A}_q$ [\cite{BSh1}, Definition 3.1]. Namely, for $k_\pm\neq 0$ one considers:
\beqa
{\cW}_+(\zeta)&\rightarrow&\sum_{k\in {\mathbb Z}_+}{\textsf W}_{-k}U(\zeta)^{-k-1} \ , \quad {\cW}_-(\zeta)\rightarrow\sum_{k\in {\mathbb Z}_+}{\textsf W}_{k+1}U(\zeta)^{-k-1} \ ,\label{c1}\\
 \quad {\cZ}_+(\zeta)&\rightarrow& \frac{1}{k_-}\sum_{k\in {\mathbb Z}_+}{\textsf G}_{k+1}U(\zeta)^{-k-1} + \frac{k_+(q+q^{-1})^2}{(q-q^{-1})} \ ,\nonumber\\
 \quad {\cZ}_-(\zeta)&\rightarrow&\frac{1}{k_+}\sum_{k\in {\mathbb Z}_+}\tilde{{\textsf G}}_{k+1}U(\zeta)^{-k-1}+ \frac{k_-(q+q^{-1})^2}{(q-q^{-1})}  \ .\nonumber
\eeqa
By analogy\footnote{Recall that the relations satisfied by the elements of ${\cal A}^{diag}_q$ follow from the ones satisfied by the elements of ${\cal A}_q$ by setting $k_\pm=0$.}, the defining relations of ${\cal A}^{diag}_q$ follow from (\ref{ec1})-(\ref{ec16}) by considering instead the homomorphism:
\beqa
{\cW}_+(\zeta)&\rightarrow&\sum_{k\in {\mathbb Z}_+}{\textsf K}_{-k}U(\zeta)^{-k-1} \ , \quad {\cW}_-(\zeta)\rightarrow\sum_{k\in {\mathbb Z}_+}{\textsf K}_{k+1}U(\zeta)^{-k-1} \ ,\label{caug1}\\
 \quad {\cZ}_+(\zeta)&\rightarrow& \sum_{k\in {\mathbb Z}_+}{\textsf Z}_{k+1}U(\zeta)^{-k-1}  \  , \quad {\cZ}_-(\zeta)\rightarrow\sum_{k\in {\mathbb Z}_+}\tilde{{\textsf Z}}_{k+1}U(\zeta)^{-k-1}\ .\nonumber
\eeqa
Strictly speaking, for $k_\pm\neq 0$ the current algebra with defining relations (\ref{ec1})-(\ref{ec16})  and (\ref{c1}) is isomorphic to ${\cal A}_q$ \cite{BSh1}. For $k_\pm=0$, the definition (\ref{caug1}) has to be considered instead: in this special case, following \cite{BSh1} the current algebra (\ref{ec1})-(\ref{ec16})  with (\ref{caug1}) is isomorphic to ${\cal A}^{diag}_q$. As both current algebras have the same defining relations and only differ by (\ref{c1}) and (\ref{caug1}), for simplicity we  keep the notation $O_q(\widehat{sl_2})$ for both cases.
\vspace{1mm}

Note that according to the results of the previous Section, the following obvious homomorphisms may be alternatively considered. There are given by \
${\cW}_\pm(\zeta)\rightarrow \overline{\cW}_\pm(\zeta)$, ${\cZ}_\pm(\zeta)\rightarrow \overline{\cZ}_\pm(\zeta)$ \
with the following substitutions in the r.h.s of the mode expansions (\ref{c1}) and (\ref{caug1}), respectively:
\beqa
&&{\textsf W}_{-k} \rightarrow \overline{\textsf W}_{-k}\ ,\quad {\textsf W}_{k+1} \rightarrow \overline{\textsf W}_{k+1}\ , \quad {\textsf G}_{k+1} \rightarrow \overline{{\textsf G}}_{k+1}\ , \quad \tilde{\textsf G}_{k+1} \rightarrow \overline{\tilde{\textsf G}}_{k+1} \ ,\quad k_\pm \rightarrow \bak_\mp \label{replacb} \\ 
&&{\textsf K}_{-k} \rightarrow \overline{\textsf K}_{-k}\ ,\quad {\textsf K}_{k+1} \rightarrow \overline{\textsf K}_{k+1}\ , \quad {\textsf Z}_{k+1} \rightarrow \overline{{\textsf Z}}_{k+1}\ , \quad \tilde{\textsf Z}_{k+1} \rightarrow \overline{\tilde{\textsf Z}}_{k+1} \ .\nonumber
\eeqa

\vspace{1mm}

As mentioned above, the half-infinite XXZ spin chain (\ref{Hsemi}) can be considered as the thermodynamic limit of (\ref{HN}). Using the results of the previous Section and the homomorphisms   (\ref{c1}), (\ref{caug1}) with (\ref{replacb}), a generating function of all mutually commuting quantities can be built, inspired by (\ref{Ifindual}) and (\ref{Ifindualaug}). We define\footnote{Note that the defining relations (\ref{ec1})-(\ref{ec16}) are invariant under the substitution $\zeta\rightarrow -\zeta^{-1}q^{-1} $}:
\beqa
&&{\ocI}(\zeta)=\epsilon_+\ocW_-(-\zeta^{-1}q^{-1}) + \epsilon_- \ocW_+(-\zeta^{-1}q^{-1})  + \frac{1}{q^2-q^{-2}}\left(k_-\ocZ_-(-\zeta^{-1}q^{-1}) + k_+\ocZ_+(-\zeta^{-1}q^{-1})\right)\ .\label{Iinf}
\eeqa

 Inspired by the Onsager's presentation of the finite chain described in the previous Section,
the transfer matrix associated with the half-infinite XXZ spin chain (\ref{Hsemi}) can now be proposed, expressed in terms of $O_q(\widehat{sl_2})$ currents acting on ${\cal V}$. By analogy with (\ref{tfin}), (\ref{tfingd}) together with (\ref{expH}) and using above quantities, in the following Sections we will consider: 
\beqa
&&t^{({\cal V})}(\zeta)= g\frac{(\zeta^2-\zeta^{-2})}{\rho(\zeta)} {\ocI}^{({\cal V})}(\zeta)\qquad \mbox{and} \qquad 
\frac{d}{d\zeta}t^{({\cal V})}(\zeta)|_{\zeta=1} = -\frac{4}{(q-q^{-1})} H_{\frac{1}{2}XXZ}\ \label{transf}
\eeqa 
where the index ${({\cal V})}$ refers to the space on which the currents act and the function $\rho(\zeta)$ is chosen such that
\beqa
t^{({\cal V})}(\zeta)=t^{({\cal V})}(-\zeta^{-1}q^{-1})\ , \quad t^{({\cal V})}(\zeta)t^{({\cal V})}(\zeta^{-1})=\id\ ,\quad  t^{({\cal V})}(\zeta)|_{\zeta=1}=\id\ .\ \nonumber
\eeqa
\vspace{1mm}

In order to study the spectral problem for (\ref{transf}), suitable infinite dimensional representations for the current algebra $O_q(\widehat{sl_2})$ need to be constructed. By analogy with the case of the infinite XXZ spin chain, recall that the fundamental operators exhibited in \cite{FM,Jim0,vertex}  are identified with Chevalley elements of $U_q(\widehat{sl_2})$  acting on an infinite tensor product of two-dimensional vector spaces, thanks to the coproduct structure. For the half-infinite XXZ spin chain (\ref{Hsemi}), the fundamental generators of ${\cal A}_q$ and ${\cal A}^{diag}_q$ can be written as linear combinations of Chevalley elements of $U_q(\widehat{sl_2})$ acting on ${\cal V}$ as we are going to show.\vspace{1mm}

We start by considering the fundamental generators of ${\cal A}_q$. Recall that the explicit expressions (\ref{op}) hold for any $N$. In the thermodynamic limit $N\rightarrow \infty$, one has:
\beqa
\qquad \qquad {\cal W}^{(\infty)}_0&=& \sum_{j=1}^{\infty}\big(  ...\otimes q^{\sigma_3} \otimes   q^{\sigma_3}\otimes \underbrace{(k_+\sigma_+ + k_-\sigma_-)}_{site j} \otimes  I\!\!I \otimes ... \otimes I\!\!I  \big)\quad + \quad \epsilon_+  \  \big(...\otimes q^{\sigma_3} \otimes q^{\sigma_3} \big)\ , \label{opfund} \\
\qquad\qquad  {\cal W}^{(\infty)}_1&=& \sum_{j=1}^{\infty}  \big( ...\otimes q^{-\sigma_3} \otimes   q^{-\sigma_3}\otimes \underbrace{(k_+\sigma_+ + k_-\sigma_-)}_{site j} \otimes  I\!\!I \otimes ... \otimes  I\!\!I  \big)\quad + \quad  \epsilon_-  \ \big(  ...\otimes q^{-\sigma_3} \otimes q^{-\sigma_3}\big)\ .\nonumber
\eeqa
Following \cite{Bas2},  one realizes the elements ${\textsf W}_0,{\textsf W}_1$ as linear combinations of Chevalley elements of  $U_q(\widehat{sl_2})$. Define
\beqa
{\textsf W}_0&=& k_+e_1 + k_-q^{-1}f_1q^{h_1} + \epsilon_+ q^{h_1} \ ,\label{realop}\\
{\textsf W}_1&=& k_-e_0 + k_+q^{-1}f_0q^{h_0} + \epsilon_-q^{h_0} \ \nonumber
\eeqa
which satisfy the defining relations of the $q-$Onsager algebra \cite{Bas2}:
\beqa \big[{\textsf W}_0,\big[{\textsf W}_0,\big[{\textsf W}_0,{\textsf
W}_1\big]_q\big]_{q^{-1}}\big]&=&\rho\big[{\textsf W}_0,{\textsf W}_1\big]\ ,\label{Talggen}\\
\big[{\textsf W}_1,\big[{\textsf W}_1,\big[{\textsf W}_1,{\textsf
W}_0\big]_q\big]_{q^{-1}}\big]&=&\rho\big[{\textsf W}_1,{\textsf W}_0\big]\ \nonumber
\eeqa
with (\ref{rho}). The fundamental operators of the half-infinite XXZ open spin chain (\ref{opfund}) are recovered as follows. For the choice\footnote{In \cite{BK1}, a different coproduct is considered.} of $U_q(\widehat{sl_2})$ coproduct considered in \cite{vertex} (see Appendix B), one introduces the coaction\footnote{In general, given a Hopf algebra ${\cal H}$ with comultiplication $\Delta$ and counit ${\cal E}$, ${\cal I}$ is
called a left ${\cal H}-$comodule (coideal subalgebra of ${\cal H}$) if there exists a 
coaction map $\delta:\ \ {\cal I}\rightarrow {\cal H}
\otimes {\cal I}$ such that (right coaction maps are defined similarly)
\beqa (\Delta \times id)\circ
\delta=(id\times\delta)\circ\delta\ ,\qquad
({\cal E} \times id)\circ \delta \cong id \
.\label{defcoaction}\nonumber\eeqa
 } map   $\delta: \quad {\cal A}_q \rightarrow U_q(\widehat{sl_2}) \otimes {\cal A}_q$  defined by:
\beqa
\delta({\textsf W}_0)&=& (k_+e_1 + k_-q^{-1}f_1q^{h_1})\otimes 1 + q^{h_1} \otimes {\textsf W}_0\ ,\label{deltadef}\\
\delta({\textsf W}_1)&=& (k_-e_0 + k_+q^{-1}f_0q^{h_0})\otimes 1 + q^{h_0} \otimes {\textsf W}_1\ .\nonumber
\eeqa
Let $\delta^{(N)}= (id \times \delta ) \circ \delta^{(N-1)}$. Then, for $N\rightarrow\infty$ it follows that $\delta^{(N)}({\textsf W}_0)$ and $\delta^{(N)}({\textsf W}_1)$ act as (\ref{opfund}) on ${\cal V}$, respectively.\vspace{1mm}

Alternatively, let us mention that another realization may be considered, that will be useful later on. Namely, the elements
\beqa
\overline{\textsf W}_0&=& \bak_+q^{-1} e_1q^{-h_1} + \bak_-f_1 + \bepsilon_- q^{-h_1} \ ,\label{orealop}\\
\overline{\textsf W}_1&=& \bak_-q^{-1} e_0q^{-h_0} + \bak_+f_0 + \bepsilon_+ q^{-h_0} \ \nonumber
\eeqa
also satisfy (\ref{Talggen}) with $\rho =(q+q^{-1})^2 \bak_+\bak_-$. In this case, for the choice of $U_q(\widehat{sl_2})$ coproduct (\ref{coprod}) the corresponding coaction map $\overline{\delta}: \quad {\cal A}_q \rightarrow {\cal A}_q \otimes U_q(\widehat{sl_2})$ is such that:  
\beqa
{\odelta}(\overline{\textsf W}_0)&=& 1 \otimes (\bak_+q^{-1} e_1q^{-h_1} + \bak_-f_1 )  +   \overline{\textsf W}_0 \otimes q^{-h_1} \ ,\label{odeltadef}\\
\odelta(\overline{\textsf W}_1)&=& 1 \otimes (\bak_-q^{-1} e_0q^{-h_0} + \bak_+f_0 ) +   \overline{\textsf W}_1 \otimes q^{-h_0} \ .\nonumber
\eeqa

We then turn to ${\cal A}^{diag}_q$. Using the explicit expressions (\ref{opdiag}), in the thermodynamic limit $N\rightarrow \infty$ the fundamental generators take the form:
\beqa
&&\qquad {\cal K}^{(\infty)}_0=  \epsilon_+\big(... \otimes q^{\sigma_3} \otimes   q^{\sigma_3}\big)  \ , \qquad {\cal K}^{(\infty)}_1=  \epsilon_-\big(... \otimes q^{-\sigma_3} \otimes   q^{-\sigma_3}\big) \  , \label{opfundaug} \\
{\cal Z}^{(\infty)}_1&=& (q^2-q^{-2})\Big(\epsilon_+\sum_{j=1}^{\infty} \big(  ...\otimes  I\!\!I \otimes \underbrace{\sigma_-}_{site j}  \otimes \,  q^{\sigma_3} \otimes   ... \otimes   q^{\sigma_3}  \big) \ + \ \epsilon_-\sum_{j=1}^{\infty} \big(  ...\otimes  I\!\!I \otimes \underbrace{\sigma_-}_{site j}  \otimes \,  q^{-\sigma_3} \otimes   ... \otimes   q^{-\sigma_3}  \big)\Big)\ , \nonumber \\
\qquad \tilde{\cal Z}^{(\infty)}_1&=&(q^2-q^{-2})\Big(\epsilon_+\sum_{j=1}^{\infty} \big(  ...\otimes  I\!\!I \otimes \underbrace{\sigma_+}_{site j}  \otimes  \,q^{\sigma_3} \otimes   ... \otimes   q^{\sigma_3} \big) \ + \  \epsilon_-\sum_{j=1}^{\infty} \big(  ...\otimes  I\!\!I \otimes \underbrace{\sigma_+}_{site j}  \otimes \, q^{-\sigma_3} \otimes   ... \otimes   q^{-\sigma_3} \big)\Big)\ . \nonumber
\eeqa
Using (\ref{realop}), it is straightforward to extract a realization in terms of $U_q(\widehat{sl_2})$ elements: indeed, note that the fundamental operators ${\cal Z}^{(\infty)}_1,\tilde{\cal Z}^{(\infty)}_1$ can be derived from the $q-$commutators $[{\cal W}^{(\infty)}_1,{\cal W}^{(\infty)}_0]_q$ and $[{\cal W}^{(\infty)}_0,{\cal W}^{(\infty)}_1]_q$, respectively, by setting $k_\pm=0$. The explicit expressions (\ref{realop}) then suggest to consider:
\beqa
{\textsf K}_0&=& \epsilon_+q^{h_1}  \ ,\qquad {\textsf K}_1= \epsilon_- q^{h_0} \ ,\label{realopaug} \\
{\textsf Z}_1&=&      (q^2-q^{-2})\big( \epsilon_+q^{-1} e_0q^{h_1}  + \epsilon_- f_1 q^{h_1+h_0}\big) \ ,\nonumber\\
\tilde{\textsf Z}_1&=& (q^2-q^{-2})\big( \epsilon_-q^{-1} e_1q^{h_0}  + \epsilon_+ f_0 q^{h_1+h_0}\big) \ \nonumber
\eeqa
which, as one can check, satisfy an augmented $q-$Onsager algebra with defining relations:
\beqa
[ {\textsf K}_0, {\textsf K}_1]&=&0\ ,\label{Tauggen}\\
 {\textsf K}_0{\textsf Z}_1&=&q^{-2} {\textsf Z}_1{\textsf K}_0\ ,\qquad {\textsf K}_0\tilde{\textsf Z}_1=q^{2}\tilde{\textsf Z}_1{\textsf K}_0\ ,\nonumber\\
{\textsf K}_1{\textsf Z}_1&=&q^{2}{\textsf Z}_1{\textsf K}_1\ ,\ \ \qquad{\textsf K}_1\tilde{\textsf Z}_1=q^{-2}\tilde{\textsf Z}_1{\textsf K}_1\ ,\nonumber\\
\big[{\textsf Z}_1,\big[{\textsf Z}_1,\big[{\textsf Z}_1,\tilde{\textsf
Z}_1\big]_q\big]_{q^{-1}}\big]&=&\rho_{diag}{\textsf Z}_1(\,{\textsf K}_1{\textsf K}_1-\,{\textsf K}_0{\textsf K}_0){\textsf Z}_1,\nonumber\\
\big[\tilde{\textsf Z}_1,\big[\tilde{\textsf Z}_1,\big[\tilde{\textsf Z}_1,{\textsf
Z}_1\big]_q\big]_{q^{-1}}\big]&=&\rho_{diag}\tilde{\textsf Z}_1({\textsf K}_0{\textsf K}_0-{\textsf K}_1{\textsf K}_1)\tilde{\textsf Z}_1\  \nonumber
\eeqa
with (\ref{rhodiag}). The coaction map that is compatible with the coproduct of $U_q(\widehat{sl_2})$  here considered as well as the relations (\ref{Tauggen}) is such that:
\beqa
\delta({\textsf K}_0)&=& q^{h_1} \otimes {\textsf K}_0\ ,\qquad \delta({\textsf K}_1)= q^{h_0} \otimes {\textsf K}_1\ ,\label{deltadefaug}\\
\delta({\textsf Z}_1)&=& q^{h_0+h_1} \otimes {\textsf Z}_1 + (q^2-q^{-2})\big( q^{-1} e_0q^{h_1} \otimes {\textsf K}_0 +  f_1q^{h_0+h_1} \otimes {\textsf K}_1\big) \ ,\nonumber\\
\delta(\tilde{\textsf Z}_1)&=&   q^{h_0+h_1} \otimes \tilde{\textsf Z}_1 + (q^2-q^{-2})\big( f_0q^{h_0+h_1} \otimes {\textsf K}_0 + q^{-1} e_1q^{h_0} \otimes {\textsf K}_1\big) \ .\nonumber
\eeqa
For $N\rightarrow\infty$ it is straightforward to check that $\delta^{(N)}({\textsf K}_0)$, $\delta^{(N)}({\textsf K}_1)$,  $\delta^{(N)}({\textsf Z}_1)$ and $\delta^{(N)}(\tilde{\textsf Z}_1)$ act as (\ref{opfundaug}) on ${\cal V}$, respectively.\vspace{1mm}

Once again, another realization of the fundamental generators of ${\cal A}^{diag}_q$ can be proposed. As one can check,
the elements 
\beqa
\overline{\textsf K}_0&=& \bepsilon_-q^{-h_1}  \ ,\qquad \overline{\textsf K}_1= \bepsilon_+ q^{-h_0} \ ,\label{orealopaug} \\
\overline{{\textsf Z}}_1&=&      (q^2-q^{-2})\big( \bepsilon_+ e_1q^{-h_0-h_1}  + \bepsilon_- q^{-1} f_0q^{-h_1}\big) \ ,\nonumber\\
\overline{\tilde{\textsf Z}}_1&=& (q^2-q^{-2})\big( \bepsilon_- e_0q^{-h_0-h_1}  + \bepsilon_+ q^{-1}f_1q^{-h_0}\big) \ \nonumber
\eeqa
also satisfy the augmented $q-$Onsager algebra (\ref{Tauggen}) and the appropriate coaction map is such that: 
\beqa
\odelta(\overline{\textsf K}_0)&=& \overline{\textsf K}_0 \otimes q^{-h_1}\ ,\qquad \odelta(\overline{\textsf K}_1)=  \overline{\textsf K}_1 \otimes q^{-h_0} \ ,\label{odeltadefaug}\\
\odelta(\overline{\textsf Z}_1)&=&  \overline{\textsf Z}_1  \otimes q^{-h_0-h_1}+ (q^2-q^{-2})\big(  \overline{\textsf K}_1 \otimes e_1q^{-h_0-h_1} +  \overline{\textsf K}_0 \otimes q^{-1} f_0q^{-h_1}\big) \ ,\nonumber\\
\odelta(\overline{\tilde{\textsf Z}}_1)&=&    \overline{\tilde{\textsf Z}}_1  \otimes q^{-h_0-h_1} +  (q^2-q^{-2})\big(  \overline{\textsf K}_0 \otimes e_0q^{-h_0-h_1} + \overline{\textsf K}_1 \otimes q^{-1}f_1q^{-h_0}\big) \ .\nonumber
\eeqa

Having identified $U_q(\widehat{sl_2})$ realizations of the fundamental generators of ${\cal A}_q$ and ${\cal A}^{diag}_q$, as well as left or right coaction maps which are compatible with the $U_q(\widehat{sl_2})$ coproduct (\ref{coprod}), we can turn to the construction of  infinite dimensional representations that will be useful in solving the spectral problem of (\ref{Hsemi}) based on the Onsager's presentation (\ref{transf}).

\section{The current algebra $O_q(\widehat{sl_2})$ and $q-$vertex operators}
The purpose of this Section is to construct  infinite dimensional representations of the current algebra (\ref{ec1}-\ref{ec16}) that will find applications in the massive regime $-1<q<0$ of the XXZ open spin chain. Besides, we will show that the $q$-vertex operators of $U_q(\widehat{sl_2})$ are intertwiners of  ${\cal A}_q$-modules or ${\cal A}^{diag}_q$-modules, giving an alternative support to the proposal of \cite{JKKKM}. As a byproduct, the $q-$boson realization of $O_q(\widehat{sl_2})$ currents recently proposed in \cite{BB} is independently confirmed.\vspace{1mm}

 Let ${\cal V}_\zeta$ be the two-dimensional evaluation representation of $U_q(\widehat{sl_2})$ in the principal picture (see Appendix B) and consider first the realization (\ref{realop}). Following \cite{vertex}, type I and type II $q$-vertex operators can be introduced such that, respectively:
\beqa
\chi(\zeta): &&\quad {\cal V} \rightarrow  {\cal V}  \otimes {\cal V}_\zeta \nonumber\  ,\\
\ochi(\zeta):&&\quad {\cal V} \rightarrow   {\cal V}_\zeta \otimes {\cal V}   \  .  \nonumber
\eeqa
According to the definition of the coaction map $\delta$ that is compatible with the realization (\ref{realop}), they satisfy (up to a scalar factor in the r.h.s):
\beqa
{\mbox Type \  I:} \qquad \chi(\zeta) \circ a &=& (id\times \pi_{\zeta} )\big[\delta(a)\big] \circ \chi(\zeta)    \label{eqvertex} \ ,\\
{\mbox Type \  II:} \qquad \ochi(\zeta)  \circ a &=& (\pi_{\zeta}\times id)\big[\delta(a)\big] \circ \ochi(\zeta)  \quad \qquad \forall a\in {\cal A}_q \ \mbox{or} \ {\cal A}^{diag}_q\ .  \nonumber
\eeqa
Writting $q$-vertex operators in the form\footnote{We set $v_+=\left(
\begin{array}{c}
  1    \\ 0
\end{array} \right)\ $ and $v_-=\left(
\begin{array}{c}
  0    \\ 1 \end{array} \right)\ $. }:
\beqa
\chi(\zeta) &=& \chi_+(\zeta) \otimes v_+   + \chi_-(\zeta) \otimes v_-\ ,\nonumber\\ 
\ochi(\zeta)&=&v_+ \otimes \ochi_+(\zeta)  + v_- \otimes \ochi_-(\zeta) \nonumber\  ,
\eeqa
two systems of equations follows from (\ref{eqvertex}). Choosing $a\equiv {\textsf W}_0,{\textsf W}_1$ and using (\ref{deltadef}), the defining relations of type II $q$-vertex operators are given by:
\beqa
{\textsf W}_0\ochi_+(\zeta) &=&   q^{-1}\ochi_+(\zeta){\textsf W}_0  -k_+ \zeta q^{-1} \ochi_-(\zeta)\ ,\label{defVOI1}\\
{\textsf W}_0\ochi_-(\zeta) &=&   q\ochi_-(\zeta){\textsf W}_0  -k_- \zeta^{-1}q \ochi_+(\zeta)\ ,\nonumber\\
{\textsf W}_1\ochi_+(\zeta) &=&   q\ochi_+(\zeta){\textsf W}_1  -k_+ \zeta^{-1}q \ochi_-(\zeta)\ ,\nonumber\\
{\textsf W}_1\ochi_-(\zeta) &=&   q^{-1}\ochi_-(\zeta){\textsf W}_1  -k_- \zeta q^{-1} \ochi_+(\zeta)\ .\label{defVOI4}\nonumber
\eeqa
For type I $q$-vertex operators, the relations (\ref{eqvertex}) hold for independent values of $k_\pm,\epsilon_\pm$. After simplications, the corresponding equations reduce to:
\beqa
&&e_0\chi_+(\zeta)=\chi_+(\zeta)e_0\ , \qquad  \qquad  \qquad \qquad  \qquad  \qquad  e_1\chi_+(\zeta) + \zeta q^{h_1}\chi_-(\zeta)=\chi_+(\zeta)e_1\ ,\label{defVOII4} \\
&&e_0\chi_-(\zeta) + \zeta q^{h_0}\chi_+(\zeta)=\chi_-(\zeta)e_0\ ,\qquad \qquad \quad  \qquad  e_1\chi_-(\zeta)=\chi_-(\zeta)e_1\ ,\nonumber \\
&&f_0q^{h_0}\chi_+(\zeta) + q\zeta^{-1}q^{h_0}\chi_-(\zeta)=\chi_+(\zeta)f_0 q^{h_0}\ ,\qquad  f_1q^{h_1}\chi_+(\zeta)=\chi_+(\zeta)f_1q^{h_1}\ ,\nonumber \\
&&f_0q^{h_0}\chi_-(\zeta)=\chi_-(\zeta)f_0q^{h_0}\ ,\qquad \qquad  \qquad  f_1q^{h_1}\chi_-(\zeta) + q\zeta^{-1}q^{h_1}\chi_+(\zeta)=\chi_-(\zeta)f_1q^{h_1}\ ,\nonumber \\
&&q^{h_0}\chi_\pm(\zeta)=q^{\pm 1}\chi_\pm(\zeta)q^{h_0}\nonumber\ ,\qquad \qquad  \qquad  q^{h_1}\chi_\pm(\zeta)=q^{\mp 1}\chi_\pm(\zeta)q^{h_1}\ ,\  \nonumber
\eeqa
which can be equally written, using the coproduct (\ref{coprod}), as:
\beqa
\chi(\zeta) \circ x &=& (id\times \pi_{\zeta} )\big[\Delta(x)\big] \circ \chi(\zeta) \quad \mbox{for} \quad x\in\{e_i,f_iq^{h_i},q^{h_i}\}\label{cochi} \ .
\eeqa
\vspace{1mm}

On the other hand, if we choose the second realization (\ref{orealop}) of ${\cal A}_q$ instead, a similar analysis can be done using the coaction map $\odelta$. It leads to
an alternative set of defining relations for type I and type II $q-$vertex operators.  
\begin{rem}
Type I $q-$vertex operators satisfy relations of the form (\ref{defVOI1}) provided the substitutions:
\beqa
\ochi_\pm(\zeta)\rightarrow \chi_\pm(\zeta^{-1}) \ ,\quad k_\pm \rightarrow \bak_\pm \ , \quad {\textsf W}_0 \rightarrow \overline{\textsf W}_{1}\ ,\quad {\textsf W}_1 \rightarrow \overline{\textsf W}_{0}\ .\label{subst1}
\eeqa 
Type II $q-$vertex operators can be defined similarly. For generic values of $\bak_\pm,\bepsilon_\pm$ the defining relations simplify to:
\beqa
\ochi(\zeta) \circ x &=& (\pi_{\zeta} \times id  )\big[\Delta(x)\big] \circ \ochi(\zeta) \quad \mbox{for} \quad x\in\{e_iq^{-h_i},f_i,q^{-h_i}\}\label{cochibar} \ .
\eeqa
\end{rem}

\vspace{2mm}

The same analysis applies to the fundamental generators of ${\cal A}^{diag}_q$. Using the realization (\ref{realopaug}) with coaction map $\delta$, two sets of equations are obtained. For instance, according to the coaction map (\ref{deltadefaug}) type II $q-$vertex operators are defined by:
\beqa
{\textsf K}_0\ochi_\pm(\zeta) &=&   q^{\mp 1}\ochi_\pm(\zeta){\textsf K}_0 \ ,\label{defVOI1aug}\\
{\textsf K}_1\ochi_\pm(\zeta) &=&   q^{\pm 1}\ochi_\pm(\zeta){\textsf K}_1 \ ,\nonumber\\
{\textsf Z}_1\ochi_+(\zeta) &=&   \ochi_+(\zeta){\textsf Z}_1 \ ,\nonumber\\
{\textsf Z}_1\ochi_-(\zeta) &=&   \ochi_-(\zeta){\textsf Z}_1 - (q^2-q^{-2})\big( \zeta q^{-1} \ochi_+(\zeta) {\textsf K}_0 + \zeta^{-1}q \ochi_+(\zeta) {\textsf K}_1 \big)
\ ,\nonumber \\
\tilde{\textsf Z}_1\ochi_+(\zeta) &=&   \ochi_+(\zeta)\tilde{\textsf Z}_1 - (q^2-q^{-2})\big( \zeta q^{-1} \ochi_-(\zeta) {\textsf K}_1 + \zeta^{-1}q  \ochi_-(\zeta) {\textsf K}_0\big)
\ ,\nonumber \\
\tilde{\textsf Z}_1\ochi_-(\zeta) &=&   \ochi_-(\zeta)\tilde{\textsf Z}_1\ .\label{defVOI4aug}\nonumber
\eeqa
The defining relations of type I $q-$vertex operators reduce to (\ref{cochi}).
\begin{rem}
Type I $q-$vertex operators satisfy relations of the form (\ref{defVOI1aug}) provided the substitutions:
\beqa
\ochi_\pm(\zeta)\rightarrow \chi_\pm(\zeta^{-1}) \ , \quad {\textsf K}_0 \rightarrow \overline{\textsf K}_{1}\ ,\quad {\textsf K}_1 \rightarrow \overline{\textsf K}_{0}\ , \quad {\textsf Z}_1 \rightarrow \overline{\tilde{\textsf Z}}_1\ , \quad \tilde{\textsf Z}_1 \rightarrow \overline{{\textsf Z}}_1 \ .\label{subst2}
\eeqa 
Type II $q-$vertex operators can be defined similarly, leading to (\ref{cochibar})
\end{rem}

More generally, for a given realization and corresponding coaction map the defining relations generalizing (\ref{defVOI1}) or (\ref{defVOI1aug}) or alternatively those associated with (\ref{subst1}) or (\ref{subst2}) - satisfied by the intertwiners for {\it any} element  of  ${\cal A}_q$ or ${\cal A}^{diag}_q$, respectively - can be obtained using the properties of the current algebra $O_q(\widehat{sl_2})$. Two different types of coaction map that generalize either $\delta$ or $\odelta$ have to be considered to this end. Namely, with minor changes in the results\footnote{Starting from a solution of the reflection equation $K(\zeta)$ in terms of the currents, it is known that $R_{12}(\zeta/v)K_{1}(\zeta)R_{12}(\zeta v)$ is also a solution of the reflection equation. This property was used in \cite{BSh1} to define a coaction map $\delta'$.} of [\cite{BSh1},Proposition 2.2], a left coaction map $\delta':{\cal A}_q\rightarrow U_q(sl_2) \otimes {\cal A}_q$ which preserves all defining relations (\ref{ec1})-(\ref{ec16}) follows. By straightforward calculations, the relations generalizing   (\ref{defVOI1}) or (\ref{defVOI1aug}) according to (\ref{c1})  and (\ref{caug1}) follow from (\ref{eqvertex}). Combining these, we eventually find that the $q-$vertex operators must satisfy:\vspace{2mm}
\beqa
\cW_-(\zeta) \ochi_-(v)&=&
\frac{\kappa(v\,\zeta)\kappa(-v\, \zeta^{-1} q^{-1})}{U(\zeta)-U(v^{-1}q)}
\Big(
( q^{-1} U(\zeta)-U(v^{-1}\sqrt{q})) \ochi_-(v)\cW_-(\zeta)
+ q\frac{q-q^{-1}}{q+q^{-1}}\ochi_-(v)\cW_+(\zeta) \label{ac}\\
&&\qquad \qquad \qquad  \qquad \qquad -\,v\,q^{-1}\frac{ \,(q-q^{-1})}{(q+q^{-1})^2 } \ochi_+(v)\cZ_-(\zeta) \Big)\nonumber\ ,\\ 
\cW_+(\zeta) \ochi_-(v)&=& 
\frac{\kappa(v\,\zeta)\kappa(-v\, \zeta^{-1} q^{-1})}{U(\zeta)-U(v^{-1}q)}
\Big( ( q \,U(\zeta)-U(\sqrt{q}\,v^{-1}))\ochi_-(v)\cW_+(\zeta) 
-q^{-1}\frac{q-q^{-1}}{q+q^{-1}}\ochi_-(v)\cW_-(\zeta)\nonumber\\
&&\qquad \qquad \qquad  \qquad \qquad -v^{-1}q\,\frac{(q-q^{-1})}{ (q+q^{-1})^2} \ochi_+(v)\cZ_-(\zeta)
\Big)\nonumber\ ,\\ 
\cZ_-(\zeta) \ochi_-(v)&=&\kappa(v\,\zeta)\kappa(-v\, \zeta^{-1} q^{-1})\Big(\ochi_-(v)\cZ_-(\zeta) \Big)\nonumber\ ,\\ 
\cZ_+(\zeta) \ochi_-(v)&=&\frac{\kappa(v\,\zeta)\kappa(-v\, \zeta^{-1} q^{-1})}{U(\zeta)-U(v^{-1}q)}\Big((U(\zeta)-U(v\,q^{-1}))\ochi_-(v)\cZ_+(\zeta)\nonumber\\ 
&& \qquad  \qquad \qquad  \qquad \qquad  \qquad- \, (q^2-q^{-2})(v\,q^{-1} \,U(\zeta)-v^{-1}q)\ochi_+(v)\cW_+(\zeta) \nonumber\\ 
&& \qquad  \qquad \qquad  \qquad \qquad  \qquad -\, (q^2-q^{-2})(v^{-1}q\,U(\zeta)-v\,q^{-1}  )\ochi_+(v)\cW_-(\zeta) \Big) \ .\nonumber
\eeqa
Changing $\cW_\pm(\zeta)\rightarrow \cW_\mp(\zeta) \ ,\  \cZ_\pm(\zeta)\rightarrow \cZ_\mp(\zeta) \ , \ochi_\pm\rightarrow \ochi_\mp$  in above formula, the action of each current on $\ochi_+(v)$ follows. Note that the prefactor in the r.h.s of (\ref{ac}) comes from the definition of the $R-$matrix (\ref{R}) with (\ref{unit}), which automatically appears in the explicit form of the coaction. As one can check, expanding the currents according to (\ref{c1}) or  (\ref{caug1}) the defining relations (\ref{defVOI1}) or (\ref{defVOI1aug}), respectively, are exactly reproduced at the leading order\footnote{Note that $\kappa(v\,\zeta)\kappa(-v\, \zeta^{-1} q^{-1})$ is invariant under the substition $\zeta \rightarrow -\zeta^{-1} q^{-1}$.} in $U(\zeta)$.\vspace{2mm}

\begin{rem} A right coaction map $\odelta':{\cal A}_q\rightarrow {\cal A}_q \otimes U_q(sl_2)$ which preserves all defining relations (\ref{ec1})-(\ref{ec16}) can be considered instead. Type I $q-$vertex operators are defined accordingly, in which case the defining relations take the form (\ref{ac}) provided the substitutions:
\beqa
\ochi_\pm(\zeta)\rightarrow \chi_\pm(\zeta^{-1})\ ,\quad \cW_\pm(\zeta) \rightarrow \overline{\cW}_\mp(\zeta) \ ,\quad \cZ_\pm(\zeta) \rightarrow \overline{\cZ}_\mp(\zeta) \ .\label{acbar}
\eeqa 
\end{rem}
Note that at the leading order in $U(\zeta)$, one recovers the defining relations (\ref{subst1}), (\ref{subst2}).\vspace{1mm}

\vspace{3mm}

The identification of the $q-$vertex operators associated with ${\cal A}_q$ and ${\cal A}^{diag}_q$ is now straightforward. On one hand, we have observed that ${\cal A}_q$ and ${\cal A}^{diag}_q$ can be both interpreted as a left (resp. right) coideal subalgebra of $U_q(\widehat{sl_2})$ using the realizations (\ref{realop}) or (\ref{realopaug}) (resp. (\ref{orealop}) or (\ref{orealopaug})).  According to the choice of realization and corresponding coaction map, we have also identified the relations satisfied by type I and type II $q-$vertex operators, namely (\ref{defVOI1}), (\ref{cochi}), (\ref{subst1}), (\ref{cochibar}), (\ref{defVOI1aug}), (\ref{defVOI4aug}) and, more generally, the relations (\ref{ac}) and (\ref{acbar}).  In particular, the relations (\ref{cochi}), (\ref{cochibar}) are nothing but the defining relations of type I and type II $q$-vertex operators of $U_q(\widehat{sl_2})$ given in Appendix C. 
Then, let $V(\Lambda_i)$, $i=0,1$, denote the integrable highest weight  level one\footnote{At level one, note that $q^{h_0+h_1}=q$.} modules  of  $U_q(\widehat{sl_2})$. Recall that type I and type II $q$-vertex operators act as:
\beqa
{\mbox Type \  I:}  \qquad \Phi^{(1-i,i)}(\zeta): &&\quad V(\Lambda_i) \rightarrow  V(\Lambda_{1-i})   \otimes {\cal V}_\zeta  \nonumber\  ,\\
{\mbox Type \  II:} \qquad {\Psi^*}^{(1-i,i)}(\zeta):&&\quad V(\Lambda_i) \rightarrow    {\cal V}_\zeta \otimes V(\Lambda_{1-i}) \   \nonumber
\eeqa
and the $q$-vertex operators can be written in the form \cite{vertex}
\beqa
\Phi^{(1-i,i)}(\zeta) &=& \Phi^{(1-i,i)}_+(\zeta) \otimes v_+  + \Phi^{(1-i,i)}_-(\zeta) \otimes v_-\ ,\nonumber\\ 
{\Psi^*}^{(1-i,i)}(\zeta)&=&v_+ \otimes {\Psi_+^*}^{(1-i,i)}(\zeta)  + v_- \otimes {\Psi_-^*}^{(1-i,i)}(\zeta) \nonumber\  .
\eeqa
Using the explicit realizations (\ref{realop}) or (\ref{realopaug}) or, alternatively (\ref{orealop}) or (\ref{orealopaug}), it is straightforward to check that the following maps
\beqa
\chi(\zeta)\rightarrow \Phi^{(1-i,i)}(\zeta) \qquad    \ochi(\zeta)\rightarrow {\Psi^*}^{(1-i,i)}(\zeta) \qquad \mbox{for any} \quad i=0,1 \label{realUqsl2}
\eeqa
provide  explicit realizations of type I and type II $q$-vertex operators of ${\cal A}_q$ and ${\cal A}^{diag}_q$. The proof solely uses the defining relations of type I and type II $q-$vertex operators of $U_q(\widehat{sl_2})$.\vspace{1mm} 

Now, an explicit expression for the $O_q(\widehat{sl_2})$ currents such that all relations (\ref{ac}) or (\ref{acbar}) are satisfied is required for completeness. To this end, recall that the  currents admit certain realizations in terms of elements satisfying a Zamolodchikov-Faddeev algebra: adapting the results of [\cite{BB}, Proposition 3.2] one shows that all defining relations (\ref{ec1})-(\ref{ec16}) are satisfied by    (\ref{realUqsl2}):
\beqa
\cW_\pm (\zeta) &\rightarrow& \frac{ \zeta q  {\Psi^*}^{(1-i,i)}_\pm(\zeta^{-1})  {\Psi^*}^{(1-i,i)}_\mp(-\zeta\,q)     +  \zeta^{-1} q^{-1} \, {\Psi^*}^{(1-i,i)}_\mp(\zeta^{-1}) {\Psi^*}^{(1-i,i)}_\pm(-\zeta\,q) }{\zeta^2q^2-\zeta^{-2}q^{-2}}\ ,\label{realW} \\
\cZ_\pm(\zeta)&\rightarrow&  (q+q^{-1}) {\Psi^*}^{(1-i,i)}_\pm(\zeta^{-1})  {\Psi^*}^{(1-i,i)}_\pm(-\zeta\,q )   \ .\nonumber
\eeqa
Indeed, type I and type II $q-$vertex operators of $U_q(\widehat{sl_2})$ satisfy the Zamolodchikov-Faddeev algebras (\ref{eqn:comI}), (\ref{eqn:psicom}), respectively (see \cite{BB} for details). Then, by straightforward calculations one checks that (\ref{realW}) exactly reproduces (\ref{ac}) using (\ref{realUqsl2}). A similar conclusion follows by substituting $\cW_\pm (\zeta)\rightarrow \ocW_\mp(\zeta)$, $\cZ_\pm (\zeta)\rightarrow \ocZ_\mp(\zeta)$ in (\ref{realW}) and using (\ref{acbar}). All these results  confirm the proposal (\ref{realUqsl2}).

\section{Eigenstates of $O_q(\widehat{sl_2})$ currents and diagonalization}
In \cite{Bas3}, two finite dimensional eigenbasis of ${\cal A}_q$ were explicitly constructed, which revealed a generalization of the property of tridiagonal pairs \cite{Ter03} for all generators of ${\cal A}_q$: as described in \cite{BK3}, in a certain basis of the truncated finite vector space ${\cal V}^{(N)}$ any generator of ${\cal A}_q$ act as a block tridiagonal matrix.  Clearly, this property extends to the generators of ${\cal A}^{diag}_q$.  Whether this property generalizes to infinite dimensional representations is an interesting question which goes beyond the scope of this article. However, based on the conjecture that the level one irreducible highest weight $U_q(\widehat{sl_2})$ representation indexed $`i'$ is embedded into the half-infinite vector space $`{\cal V}'$ using $q-$vertex operators \cite{JKKKM}, for $-1<q<0$ it is possible to construct an analog of the two eigenbasis discussed in \cite{Ter03} for  $O_q(\widehat{sl_2})$. Using previous results, for the discussion below we focus on the spectral problem associated with the currents:
\beqa
&&\ocW_\pm^{(i)}(\zeta)\ , \  \ocZ_\pm^{(i)}(\zeta): \quad V(\Lambda_i) \rightarrow  V(\Lambda_{i})   \qquad \mbox{for} \qquad i=0,1 \nonumber 
\eeqa
with
\beqa
\ocW_\pm^{(i)}(\zeta) &=& \frac{ \zeta q \Phi_\mp^{(i,1-i)}(\zeta) \Phi_\pm^{(1-i,i)}(-\zeta^{-1}q^{-1})     +  \zeta^{-1}q^{-1} \,\Phi_\pm^{(i,1-i)}(  \zeta)\Phi_\mp^{(1-i,i)}(-\zeta^{-1}q^{-1}) }{\zeta^2q^2-\zeta^{-2}q^{-2}}\ ,\label{realWi} \\
\ocZ_\pm^{(i)}(\zeta)&=&   (q+q^{-1})\Phi_\mp^{(i,1-i)}(\zeta) \Phi_\mp^{(1-i,i)}(-\zeta^{-1}q^{-1})   \ .\label{realZi} \ 
\eeqa

On one hand, consider the spectral problem  
\beqa
\ocW_\pm^{(i)}(-\zeta^{-1}q^{-1}) |B_{\pm}\rangle = \lambda_{\pm}(-\zeta^{-1}q^{-1}) |B_{\pm}\rangle  \qquad \mbox{for} \qquad  i=0,1\ .
\eeqa
Acting with $g\Phi_\mp(\zeta^{-1})$ (or, alternatively $g\Phi_\pm(\zeta^{-1})$) from the left on this equation and using the properties of $q-$vertex operators (see Appendix C), it yields to:
\beqa
\frac{\zeta^{\pm 1}}{\zeta^2-\zeta^{-2}} \Phi_\mp^{(1-i,i)}(\zeta)  |B_{+}\rangle  = g  \lambda_{+}(-\zeta^{-1}q^{-1})  \Phi_\mp^{(1-i,i)}(\zeta^{-1}) |B_{+}\rangle  \ ,\label{eq1}\\
\frac{ \zeta^{\mp 1}}{\zeta^2-\zeta^{-2}} \Phi_\mp^{(1-i,i)}(\zeta)  |B_{-}\rangle  = g  \lambda_{-}(-\zeta^{-1}q^{-1})  \Phi_\mp^{(1-i,i)}(\zeta^{-1}) |B_{-}\rangle  \ .\label{eq2}
\eeqa
For each current, by straightforward calculations one finds the `minimal' solution:
\beqa
&&|B_+\rangle = e^{\tilde{F}_0}|0\rangle \qquad \mbox{and} \qquad  |B_-\rangle = e^{\alpha/2}e^{\tilde{F}_0}|0\rangle \label{Bpm}
\eeqa
where
\beqa
\tilde{F}_0=-\frac{1}{2}\sum_{n=1}^{\infty} \frac{n q^{6n}}{[2n][n]}a^2_{-n} - \sum_{n=1}^{\infty}   \frac{q^{5n/2}(1-q^n)\theta_n}{[2n]}  a_{-n}\  \quad \mbox{with}\quad \theta_n=1(0)\ \ \mbox{for}\ \ n \ \mbox{even (odd)}\ .\nonumber
\eeqa
Accordingly, the spectrum reads:
\beqa
\lambda_+(\zeta) = \lambda_-(\zeta) = \frac{1}{g}\frac{\zeta^{-1}q^{-1}}{\zeta^2q^2-\zeta^{-2}q^{-2}} \frac{\delta(\zeta^2q^2)}{\delta(\zeta^{-2}q^{-2})} \quad \mbox{where} \quad
\delta(z)= \frac{(q^6z^2;q^8)_\infty}{(q^8z^2;q^8)_\infty}\ .
\eeqa
\begin{rem} The eigenvector $|B_{+}\rangle$ (resp. $|B_{-}\rangle$) coincides exactly with $|0\rangle_B|_{r\rightarrow 0}$ (resp. $e^{\alpha/2}|1\rangle_B|_{r\rightarrow \infty}$) in \cite{JKKKM}. Also, the eigenvectors $|B_{\pm}\rangle$ do not depend on the index $i=0,1$. As a consequence, $\ocW_\pm^{(i)}(\zeta) |B_{\pm}\rangle = \ocW_\pm^{(1-i)}(\zeta) |B_{\pm}\rangle $.
\end{rem}

More generally, two families of eigenstates of $\ocW_\pm^{(i)}(\zeta)$ can be constructed using the properties of type II $q-$vertex operators that are reported in Appendix C: starting from $|B_{\pm}\rangle$, for any $ i=0,1$ the commutation relations (\ref{eqn:phipsi}) imply:
\beqa
\ocW_\pm^{(i)}(-\zeta^{-1}q^{-1}) \Psi^*_{\mu_1}(\xi_1)...\Psi^*_{\mu_m}(\xi_m)|B_{\pm}\rangle = \lambda_{\pm}(-\zeta^{-1}q^{-1};\xi_1,...,\xi_m)\Psi^*_{\mu_1}(\xi_1)...\Psi^*_{\mu_m}(\xi_m) |B_{\pm}\rangle \ 
\eeqa
with
\beqa
\lambda_\pm(\zeta;\xi_1,...,\xi_m) =  \prod_{j=1}^{m}\tau(\zeta/\xi_j)\tau(\zeta\xi_j)\lambda_\pm(\zeta) \ .\nonumber
\eeqa

The action of other currents for any $ i=0,1$  immediately follows:
\beqa
\ocZ_\epsilon^{(i)}(-\zeta^{-1}q^{-1}) \Psi^*_{\mu_1}(\xi_1)...\Psi^*_{\mu_m}(\xi_m)|B_{\epsilon'}\rangle \!\!\!\!\!\!\!\!\!\!\!\!\!\!\!\!\!\!\!\!\!&=&\nonumber \\
 (q+q^{-1}) \zeta^{1-\epsilon\epsilon'} \frac{\delta(\zeta^2q^2)}{\delta(\zeta^{-2}q^{-2})}\!\!\! &\prod_{j=1}^{m}\tau(\zeta/\xi_j)\tau(\zeta\xi_j)&\!\!\!{\Phi^*_\epsilon}^{(i,1-i)}(\zeta^{-1})\Phi_{-\epsilon}^{(1-i,i)}(\zeta^{-1}) \Psi^*_{\mu_1}(\xi_1)...\Psi^*_{\mu_m}(\xi_m) |B_{\epsilon'}\rangle \ .\nonumber
\eeqa
\vspace{1mm}

Having explicit expressions for  $\ocW_\pm^{(i)}(\zeta)$ currents' eigenstates, following \cite{BK3} we now turn to the diagonalization of the Hamiltonian (\ref{Hsemi}) in the massive regime $-1<q<0$ within Onsager's approach. 
In Section 2, we have introduced the corresponding transfer matrix in terms of the generating function of all mutually commuting quantities
(\ref{Iinf}) that form the so-called $q-$Dolan-Grady hierarchy, by analogy with the finite size case. According to (\ref{ec1})-(\ref{ec16}), the commutation and invertibility relations of the $q-$vertex operators (see Appendix C) and (\ref{realWi}), (\ref{realZi}), observe that the following relations are satisfied:
\beqa
&&\big[{\ocI}^{(i)}(\zeta),{\ocI}^{(i)}(\xi)\big]=0\ ,\quad  
{\ocI}^{(i)}(\zeta)=\kappa(-q\zeta^2){\ocI}^{(i)}(-\zeta^{-1}q^{-1})\ ,\quad g(\zeta^2-\zeta^{-2}){\ocI}^{(i)}(\zeta)|_{\zeta=1}= \epsilon_+ + \epsilon_-\ ,\nonumber\\
&&-g^2(\zeta^2-\zeta^{-2})^2{\ocI}^{(i)}(\zeta){\ocI}^{(i)}(\zeta^{-1}) =  (\epsilon_+ + \epsilon_-)^2 + (\zeta-\zeta^{-1})^2 \epsilon_+ \epsilon_- - \frac{(\zeta^2-\zeta^{-2})^2}{(q-q^{-1})^2}k_+k_- \ .\nonumber 
\eeqa
As a consequence, the following contraints on the normalization factor in terms of the boundary parameters follow:
\beqa
\label{constr}\frac{\rho(\zeta)}{\rho(-q^{-1}\zeta^{-1})} &=& -\frac{1}{\kappa(-q\zeta^2)}\frac{(\zeta^2-\zeta^{-2})}{(q^{2}\zeta^2-q^{-2}\zeta^{-2})}\ \ ,\\
\rho(\zeta)\rho(\zeta^{-1})&=& (\epsilon_+ + \epsilon_-)^2 + (\zeta-\zeta^{-1})^2\epsilon_+\epsilon_- - \frac{(\zeta^2-\zeta^{-2})^2}{(q-q^{-1})^2}k_+k_- \ ,\nonumber\\
\rho(1)&=& \epsilon_+ + \epsilon_-\ .\nonumber
\eeqa
Note that  if one writes (\ref{tfin}) solely in terms of $q-$vertex operators, one recovers exactly the transfer matrix proposed in [\cite{JKKKM}, eq. (2.13)]. In this case, the scalar solution (\ref{K-}) of the reflection equation associated with the right boundary can be explicitly exhibited.\vspace{1mm}

We are now in position to study the spectral problem associated with (\ref{Iinf}) for any choice of boundary parameters $\epsilon_\pm,k_\pm$. In the present article, for simplicity we will focus on two special cases: first, we will study the case of {\it diagonal} boundary conditions $\epsilon_\pm\neq 0,k_\pm=0$: it has been considered in details in the literature \cite{JKKKM} and will serve as a check of the approach here presented. Then, we will consider the case of upper ($\epsilon_\pm\neq 0$, $k_+\neq 0$, $k_-=0$)  or lower ($\epsilon_\pm\neq 0$, $k_-\neq 0$, $k_+=0$)  non-diagonal boundary conditions: the results here obtained can be compared with the known ones obtained within the Bethe ansatz approach \cite{CRS}. Note that the explicit expression for the vacuum vectors for upper or lower non-diagonal boundary conditions allows to derive an integral representation for correlation fonctions, following \cite{vertex,JKKKM}. This will be considered elsewhere.  

\subsection{The diagonal case revisited}
For {\it diagonal} boundary conditions $k_\pm=0$ and $\epsilon_\pm\neq 0$, define $v^2\equiv r=-\epsilon_+/\epsilon_-$. By straightforward calculations, the solution $\rho(\zeta)$ to the constraints (\ref{constr}) that is compatible with the action of the $q-$vertex operators\footnote{Indeed, another solution to (\ref{constr}) may be considered.} is given by:
\beqa
\rho(\zeta)= (\zeta\epsilon_- + \zeta^{-1}\epsilon_+)\frac{\delta(\zeta^{-2})}{\delta(\zeta^2)}\frac{\tilde{\varphi}(\zeta^{-2};r)}{\tilde{\varphi}(\zeta^{2};r)}\ \label{normaldiag}
\eeqa
where
\beqa
\tilde{\varphi}(z;r)=\frac{(q^4rz;q^4)_\infty}{(q^2rz;q^4)_\infty}\  \quad \mbox{and} \quad \delta(z)= \frac{(q^6z^2;q^8)_\infty}{(q^8z^2;q^8)_\infty}\ .\nonumber
\eeqa

Now, starting from the eigenstates $|B_\pm\rangle$ of ${\ocW}^{(i)}_\pm(\zeta)$ defined by (\ref{Bpm}), we are looking for the vacuum vectors of the transfer matrix (\ref{transf}) for $k_\pm=0$ .\vspace{2mm}

\underline{First vacuum vector $|0\rangle_B$}: Define
\beqa
|0\rangle_B \equiv e^{f(v)}|B_+\rangle\quad  \mbox{where}\quad  f(v)=-\sum_{n=1}^\infty {\frac{a_{-n}}{[2n]}}q^{7n/2}v^{2n}\nonumber\ .
\eeqa
The action of the $q-$vertex operators on the exponential term is such that:
\beqa
\Phi^{(1-i,i)}_-(\zeta) e^{f(v)} &=&     \tilde{\varphi}(\zeta^{-2};v^2)  e^{f(v)}\Phi^{(1-i,i)}_-(\zeta)\ ,\nonumber\\
\Phi^{(1-i,i)}_+(\zeta) e^{f(v)} &=&   \tilde{\varphi}(\zeta^{-2};v^2) \left((1-v^2\zeta^{-2})\  e^{f(v)}\Phi^{(1-i,i)}_+(\zeta) +  v^2(1-q^2)\zeta^{-1} e^{f(v)}\Phi^{(1-i,i)}_-(\zeta)x^{-}_{-1} \right) \ ,\nonumber
\eeqa
where we introduced the Drinfeld's generator (\ref{eqn:xpm})
\beqa 
x^{-}_{-1}= \oint_{C_1} {dw \over 2\pi i}X^{-}(w)\ .\nonumber
\eeqa
Using (\ref{eq1}), (\ref{eq2}) and noticing that  $x^{-}_{-1}|B_+\rangle = 0$ by straightforward calculations, one derives 
\beqa
\Phi_-^{(1,0)}(\zeta)  \  e^{f(v)}|B_{+}\rangle  &=& \frac{\delta(\zeta^{-2})\tilde{\varphi}(\zeta^{-2};v^2)}{\delta(\zeta^{2})\tilde{\varphi}(\zeta^{2};v^2)}\  \Phi_-^{(1,0)}(\zeta^{-1}) \  e^{f(v)}|B_{+}\rangle  \ ,\label{eqdiag1}\\
\Phi_+^{(1,0)}(\zeta)  \  e^{f(v)}|B_{+}\rangle  &=&\frac{(\zeta^2-v^2)}{(1-v^2\zeta^{2})} \frac{\delta(\zeta^{-2})\tilde{\varphi}(\zeta^{-2};v^2)}{\delta(\zeta^{2})\tilde{\varphi}(\zeta^{2};v^2)} \ \Phi_+^{(1,0)}(\zeta^{-1})\  e^{f(v)} |B_{+}\rangle  \ .\label{eqdiag2}
\eeqa
Note that this result is in agreement with \cite{JKKKM}, although the notations here differ. As a consequence, the action of the currents on the state $|0\rangle_B$ is given by:
\beqa
{\ocW}^{(0)}_+(-q^{-1}\zeta^{-1})|0\rangle_B &=& \frac{\varphi(\zeta^{-2};r)}{\varphi(\zeta^{2};r)}\left(\frac{1}{g}\frac{(\zeta-r\zeta^{-1})}{(\zeta^2-\zeta^{-2})(1-r\zeta^2)} - \frac{r\zeta}{1-r\zeta^2} {\Phi^*_-}^{(0,1)}(\zeta^{-1})\Phi_-^{(1,0)}(\zeta^{-1})\right)|0\rangle_B \nonumber \ ,\\
{\ocW}^{(0)}_-(-q^{-1}\zeta^{-1})|0\rangle_B &=& \frac{\varphi(\zeta^{-2};r)}{\varphi(\zeta^{2};r)}\left(\frac{1}{g}\frac{\zeta^2(\zeta-r\zeta^{-1})}{(\zeta^2-\zeta^{-2})(1-r\zeta^2)} - \frac{\zeta}{1-r\zeta^2} {\Phi_-^*}^{(0,1)}(\zeta^{-1})\Phi_-^{(1,0)}(\zeta^{-1})\right)|0\rangle_B \nonumber \ 
\eeqa
where the notation $\varphi(z;r)=\delta(z)\tilde{\varphi}(z;r)$ has been introduced to fit with \cite{JKKKM}. Combining both expressions together according to (\ref{Iinf}), the off-diagonal contribution cancels.  Using (\ref{normaldiag}), in agreement with \cite{JKKKM} one finds:
\beqa
t^{(0)}(\zeta)|_{k_\pm=0}|0\rangle_B = 1 \ |0\rangle_B \ .
\eeqa

\vspace{3mm}

\underline{Second vacuum vector $|1\rangle_B$}: Similar analysis can be done for the second eigenstate, denoted $|1\rangle_B$ in \cite{JKKKM}. Define
\beqa
|1\rangle_B \equiv       e^{-f(-q^{-1}v^{-1})} |B_-\rangle \ .\nonumber 
\eeqa
By straightforward calculations, one finds that 
\beqa
\qquad &&\Phi_-^{(0,1)}(\zeta)  \  e^{-f(-q^{-1}v^{-1})}|B_{-}\rangle  = \Lambda(\zeta;v^2)\frac{\delta(\zeta^{-2})\tilde{\varphi}(\zeta^{-2};v^2)}{\delta(\zeta^{2})\tilde{\varphi}(\zeta^{2};v^2)}\  \Phi_-^{(0,1)}(\zeta^{-1}) \  e^{-f(-q^{-1}v^{-1})}|B_{-}\rangle  \ ,\label{eqdiagp1}\\
\qquad &&\Phi_+^{(0,1)}(\zeta)  \  e^{-f(-q^{-1}v^{-1})}|B_{-}\rangle  = \Lambda(\zeta;v^2) \frac{(\zeta^2-v^2)}{(1-v^2\zeta^{2})} \frac{\delta(\zeta^{-2})\tilde{\varphi}(\zeta^{-2};v^2)}{\delta(\zeta^{2})\tilde{\varphi}(\zeta^{2};v^2)} \ \Phi_+^{(0,1)}(\zeta^{-1})\  e^{-f(-q^{-1}v^{-1})} |B_{-}\rangle  \ \label{eqdiagp2}
\eeqa
where
\beqa
\Lambda(\zeta;v^2) = \zeta^2\frac{\tilde{\varphi}(q^{-2}\zeta^{2};v^{-2})  \tilde{\varphi}(\zeta^{2};v^2)}{ \tilde{\varphi}(q^{-2}\zeta^{-2};v^{-2}) \tilde{\varphi}(\zeta^{-2};v^{2})} \ .\nonumber
\eeqa
The action of the currents ${\ocW}^{(1)}_\pm(-q^{-1}\zeta^{-1})$ on  $|1\rangle_B$ follows, which leads to 
\beqa
t^{(1)}(\zeta)|_{k_\pm=0}|1\rangle_B = \Lambda(\zeta;r) \ |1\rangle_B \ \nonumber
\eeqa
in agreement with \cite{JKKKM}. Finally, according to the observation that (see Appendix C and (\ref{realW})): 
\beqa
\ocW_\pm^{(i)}(\zeta) \Psi^{*(i,1-i)}_{\mu}(\xi) = \tau(\zeta/\xi)\tau(\zeta\xi) \ \Psi^{*(i,1-i)}_{\mu}(\xi) \ocW_\pm^{(i)}(\zeta) \ ,\nonumber
\eeqa
more general eigenstates of the transfer matrix are generated using type II $q-$vertex operators. In agreement with \cite{JKKKM}, it follows:
\beqa
t^{(i)}(\zeta)|_{k_\pm=0}\ \Psi^*_{\mu_1}(\xi_1)...\Psi^*_{\mu_m}(\xi_m)|i\rangle_B =\Lambda^{(i)}(\zeta;r)  \prod_{j=1}^{m}\tau(\zeta/\xi_j)\tau(\zeta\xi_j)\ \Psi^*_{\mu_1}(\xi_1)...\Psi^*_{\mu_m}(\xi_m)|i\rangle_B \ .\nonumber
\eeqa
where $\Lambda^{(0)}(\zeta;r)=1$  and $\Lambda^{(1)}(\zeta;r)=\Lambda(\zeta;r)$. From (\ref{transf}), the energy levels are derived and expressed in terms of Jacobi elliptic functions \cite{JKKKM}.

\vspace{3mm}

\subsection{Upper or lower non-diagonal boundary conditions}
Let us consider the Hamiltonian (\ref{Hsemi}) with upper {\it non-diagonal} boundary conditions $\epsilon_\pm\neq 0,k_+\neq 0$ and $k_-=0$, keeping above  parametrization $v^2\equiv r=-\epsilon_+/\epsilon_-$. Clearly, in this case the solution $\rho(\zeta)$ is also given by (\ref{normaldiag}) as the product of non-diagonal parameters $k_+k_-$ vanishes.
According to the results of Section 3, let us consider by analogy with (\ref{orealop}) the following realization of the $q-$Onsager algebra:
\beqa
\overline{\textsf w}_0&=& k'_+q^{-1} e_1q^{-h_1} + k'_-f_1  \qquad \mbox{and}\qquad \overline{\textsf w}_1= k''_-q^{-1} e_0q^{-h_0} + k''_+f_0  \ ,\label{elqons}
\eeqa
where the parameters $k'_\pm,k''_\pm$ are not determined yet. Their action on type I $q-$vertex operators are deduced from (\ref{defVOII4}). Now, define:
\beqa
\overline{\textsf w}^{(\pm)}_0 \equiv \overline{\textsf w}_0|_{k'_\mp=0}  \qquad \mbox{and} \qquad  \overline{\textsf w}^{(\pm)}_1 \equiv \overline{\textsf w}_1|_{k''_\mp=0}  \nonumber \ .
\eeqa
By straightforward calculation, it follows:
\beqa
\Phi_-(\zeta)(\overline{\textsf w}^{(+)}_0)^n &=&   q^n (\overline{\textsf w}^{(+)}_0)^n\Phi_-(\zeta)\ ,\nonumber\\
\Phi_+(\zeta)(\overline{\textsf w}^{(+)}_0)^n &=&   q^{-n} (\overline{\textsf w}^{(+)}_0)^n\Phi_+(\zeta) + k'_+ \zeta [n]_q (\overline{\textsf w}^{(+)}_0)^{n-1} \Phi_-(\zeta) \ ,\nonumber\\
\Phi_-(\zeta) (\overline{\textsf w}^{(-)}_0)^n &=&    q^{n} (\overline{\textsf w}^{(-)}_0)^n\Phi_-(\zeta) + k'_- \zeta^{-1} [n]_q  (\overline{\textsf w}^{(-)}_0)^{n-1} \Phi_+(\zeta)\ ,\nonumber\\
\Phi_+(\zeta) (\overline{\textsf w}^{(-)}_0)^n &=&   q^{-n}  (\overline{\textsf w}^{(-)}_0)^n\Phi_+(\zeta) \ \nonumber
\eeqa
and similarly for $\overline{\textsf w}^{(\pm)}_1$, provided the substitutions $\overline{\textsf w}^{(\pm)}_0 \rightarrow \overline{\textsf w}^{(\pm)}_1 $, $k'_\pm\rightarrow k''_\pm$, $\zeta\rightarrow \zeta^{-1}$ and $q\rightarrow q^{-1}$ in above commutation relations. According to (\ref{eqdiag1}),  (\ref{eqdiag2}),  (\ref{eqdiagp1}),  (\ref{eqdiagp2}), let us consider the following combinations:  
\beqa
|+;0\rangle = \sum_{n=0}^{\infty} \frac{q^{-n(n-1)/2}}{[n]_q!} (\overline{\textsf w}^{(+)}_1)^n |0\rangle_B \qquad \mbox{and} \qquad |+;1\rangle = \sum_{n=0}^{\infty} \frac{q^{n(n-1)/2}}{[n]_q!} (\overline{\textsf w}^{(+)}_0)^n |1\rangle_B \ .\label{vacND+}
\eeqa
Acting with type I $q-$vertex operators,  it is easy to show that:
\beqa
\Phi_-^{(1,0)}(\zeta)  \  |+;0\rangle &=&   \frac{\varphi(\zeta^{-2};r)}{\varphi(\zeta^{2};r)} \  \Phi_-^{(1,0)}(\zeta^{-1}) \  |+;0\rangle \ ,\label{eqdiagpri1}\\
\Phi_+^{(1,0)}(\zeta)  \  |+;0\rangle  &=& \frac{\varphi(\zeta^{-2};r)}{\varphi(\zeta^{2};r)}\ \left(\frac{(\zeta^2-v^2)}{(1-v^2\zeta^{2})} \Phi_+^{(1,0)}(\zeta^{-1}) \ - k''_+ \frac{\zeta(\zeta^2-\zeta^{-2})}{1-v^2\zeta^2}\ \Phi_-^{(1,0)}(\zeta^{-1}) \  \right)\  |+;0\rangle  \ \label{eqdiagpri2}
\eeqa
and
\beqa
&&\Phi_-^{(0,1)}(\zeta)  \  |+;1\rangle =  \Lambda(\zeta;v^2) \frac{\varphi(\zeta^{-2};r)}{\varphi(\zeta^{2};r)} \  \Phi_-^{(0,1)}(\zeta^{-1}) \  |+;1\rangle \ ,\label{eqdiagpri3}\\
&&\Phi_+^{(0,1)}(\zeta)  \  |+;1\rangle  = \Lambda(\zeta;v^2) \frac{\varphi(\zeta^{-2};r)}{\varphi(\zeta^{2};r)}\ \left(\frac{(\zeta^2-v^2)}{(1-v^2\zeta^{2})} \Phi_+^{(0,1)}(\zeta^{-1}) \ - k'_+ \frac{v^2\zeta(\zeta^2-\zeta^{-2})}{1-v^2\zeta^2}\ \Phi_-^{(0,1)}(\zeta^{-1}) \  \right)\  |+;1\rangle  \ .\label{eqdiagpri4}\
\eeqa

\vspace{2mm}

By analogy with the case of diagonal boundary conditions, it is straigthforward to derive the action of the conserved currents (\ref{Iinf}) for $k_-=0$ on above states. Appart from terms that already appeared in the case of diagonal boundary conditions, the structure of the states $|+;i\rangle$ generates an additional contribution associated with the currents ${\ocW}^{(i)}_\pm(\zeta)$ which mixes with the one associated with ${\ocZ}^{(i)}_+(\zeta)$, $i=0,1$.
Assuming that $|+;i\rangle$ are eigenstates of (\ref{Iinf}) for $k_-=0$ determines uniquely the choice of parameters $k'_+,k''_+$. Namely,
\beqa
\overline{\cal I}^{(0)}(\zeta)|_{k_-=0}|+;0\rangle &=&  \Lambda^{(0)}(\zeta^2,r)\frac{\varphi(\zeta^{-2},r)}{\varphi(\zeta^{2},r)} \frac{(\epsilon_+\zeta^{-1}+\epsilon_- \zeta ) }{g(\zeta^2-\zeta^{-2})}|+;0\rangle \ ,\nonumber\\
\overline{\cal I}^{(1)}(\zeta)|_{k_-=0}|+;1\rangle &=&  \Lambda^{(1)}(\zeta^2,r)\frac{\varphi(\zeta^{-2},r)}{\varphi(\zeta^{2},r)} \frac{(\epsilon_+\zeta^{-1}+\epsilon_- \zeta ) }{g(\zeta^2-\zeta^{-2})}|+;1\rangle \nonumber\ 
\eeqa
for
\beqa
k''_+=\frac{1}{q-q^{-1}}\frac{k_+}{\epsilon_-} \qquad \mbox{and} \qquad k'_+=-\frac{1}{q-q^{-1}}\frac{k_+}{\epsilon_+}\label{condk+}\ .
\eeqa
Although the vacuum vectors (\ref{vacND+}) with (\ref{condk+}) are more complicated than in the diagonal case, the spectrum of the transfer matrix is clearly unchanged. Note that such phenomena is known for the {\it finite} size open XXX spin chain with upper or lower non-diagonal boundary conditions \cite{CRS} (see also \cite{MMR}) within the Bethe ansatz framework. Having identified the vacuum vectors, excited states follow using the action of type II $q-$vertex operators.
\vspace{1mm}

For completeness, let us finally describe the  vacuum eigenstates of the Hamiltonian (\ref{Hsemi}) for lower non-diagonal boundary conditions  $\epsilon_\pm\neq 0,k_-\neq 0$ and $k_+=0$. They are given by:
\beqa
|-;0\rangle = \sum_{n=0}^{\infty} \frac{q^{n(n-1)/2}}{[n]_q!} (\overline{\textsf w}^{(-)}_1)^n |0\rangle_B \qquad \mbox{and} \qquad |-;1\rangle = \sum_{n=0}^{\infty} \frac{q^{-n(n-1)/2}}{[n]_q!} (\overline{\textsf w}^{(-)}_0)^n |1\rangle_B \ \label{vacND-}
\eeqa
where 
\beqa
k''_-=-\frac{1}{q-q^{-1}}\frac{k_-}{\epsilon_-} \qquad \mbox{and} \qquad k'_-=\frac{1}{q-q^{-1}}\frac{k_-}{\epsilon_+}\label{condk-}\ .
\eeqa
Note that the vacuum vectors (\ref{vacND+}) with (\ref{condk+}) and (\ref{vacND-}) with (\ref{condk-}) are power series in $k_+$ and $k_-$, respectively. For the special case $k_+=0$ in (\ref{vacND+}) (or $k_-=0$ in (\ref{vacND-})), all terms in the series disappear except the term $n=0$, in which case the vacuum vectors reduce to the ones associated with diagonal boundary conditions $k_\pm=0$. 
\vspace{2mm}

\section{Comments and perspectives}
In the present article, the research program initiated in \cite{Bas1,Bas2} and further explored in \cite{BK,BK1,BK3} has been applied to the thermodynamic limit of the XXZ open spin chain with general integrable boundary conditions. It has been shown that the formulation of the finite size case of \cite{BK3} - here completed for diagonal boundary conditions - can be directly extended to the infinite limit. In this approach, the new current algebra introduced in \cite{BSh1} plays a central role: the diagonalisation of the Hamiltonian/transfer matrix is reduced to the study of the representation theory of the current algebra $O_q(\widehat{sl_2})$. For $-1<q<0$, explicit realizations of the currents in terms of $U_q(\widehat{sl_2})$ $q-$vertex operators have been obtained, confirming independently the proposal of \cite{BB}. Also, certain properties reminiscent of the finite size case - for instance, the existence of two `dual' families of eigenstates - have been exhibited. For diagonal boundary conditions, the spectrum and eigenstates have been described in details within the new framework, providing a fresh look at the known results in \cite{JKKKM}. For upper or lower non-diagonal boundary, for the first time the spectrum and eigenstates are obtained explicitly. In particular, the eigenstates are generated starting from the currents' eigenstates through the action of Chevalley elements of $U_q(\widehat{sl_2})$. Importantly, this result and its possible generalization to generic boundary conditions - see some comments below - open the possibility to derive integral representations of correlation functions and form factors for non-diagonal boundary conditions.
\vspace{1mm} 

For generic boundary conditions, the spectral problem could be considered along the same line: equations extending (\ref{eqdiagpri1})-(\ref{eqdiagpri4}) have to be considered, where, roughly speaking,  the eigenstates are such that an additional term is generated in the r.h.s. of (\ref{eqdiagpri1}) and (\ref{eqdiagpri3}). Actually, in view of the fact that the elements (\ref{elqons}) generate a $q-$Onsager algebra, according to the intertwining relations of the form (\ref{defVOI1}) it is thus natural to construct the eigenstates of (\ref{Hsemi}) using linearly independent monomials in $\overline{\textsf w}_0,\overline{\textsf w}_1$. Namely, understanding further the construction of a Poincar\'e-Birkhoff-Witt basis of the $q-$Onsager algebra is highly desirable. In this direction, the results of \cite{BB3} suggest to consider the following combinations of `descendents' acting on the `diagonal' vacuum vectors: 
\beqa
\cW^{\alpha_1}_{-k_1}...\cW^{\alpha_N}_{-k_N}    \cG^{\beta_1}_{p_1+1}...\cG^{\beta_P}_{p_P+1}    \cW^{\gamma_M}_{l_M+1}...\cW^{\gamma_1}_{l_1+1}|i\rangle_{B} \label{basis1} 
\eeqa
where $\{\alpha_j,\beta_j,\gamma_j,k_j,p_j,l_j\}\in{\mathbb Z}_+$ and the ordering $k_1<...<k_N$; $\l_1<...<l_M$; $p_1<...<p_P$ is chosen.
We intend to study this problem separately.\vspace{1mm}

As the reader noticed, the long standing question of the non-Abelian symmetry of the Hamiltonian (\ref{Hsemi}) has not been adressed up to now although it played a central role in the initial development of the vertex operator program \cite{vertex,JKKKM}: in the case of the infinite XXZ spin chain, recall that the $U_q(\widehat{sl_2})$ algebra emerges as a non-Abelian symmetry of the Hamiltonian \cite{FM}. For generic {\it non-diagonal} or {\it diagonal} boundary conditions, following \cite{FM,Jim0} it is easy to show that a similar phenomena occurs in the thermodynamic limit of the open spin chain. Let $H_{\frac{1}{2}XXZ}=H_0+h_b$ where $H_0$ and $h_b$ denote the bulk and boundary contributions in the Hamiltonian (\ref{Hsemi}) for $\epsilon_\pm\neq0$, $k_\pm\neq 0$, respectively. By straightforward calculations one finds:
\beqa
 \big[H_0,{\cal W}^{(\infty)}_0\big]&=& -\big[ h_b,{\cal W}^{(\infty)}_0\big]=-\frac{1}{2}(q-q^{-1}) \big(...\otimes q^{\sigma_3} \otimes   q^{\sigma_3}\otimes (k_+\sigma_+ - k_-\sigma_-)\big)\ .\nonumber
\eeqa
A similar analysis can be done for  ${\cal W}^{(\infty)}_1$. Combining both expressions, one finally shows that the Hamiltonian (\ref{Hsemi}) is commuting with these operators:
\beqa
&&\big[H_{\frac{1}{2}XXZ}, a\big] = 0 \ ,\qquad \ a\in \{{\cal W}^{(\infty)}_0,{\cal W}^{(\infty)}_1\}\ . \label{sym}
\eeqa  

Similarly, for the case of generic {\it diagonal} boundary conditions, in the thermodynamic limit $N\rightarrow \infty$ by straightforward calculations one finds that the contributions coming from the commutator of the bulk and boundary terms  of the Hamiltonian with the fundamental elements of ${\cal A}^{diag}_q$ cancel each other:  
\beqa
\big[H^{diag}_{\frac{1}{2}XXZ}, a\big] = 0 \ ,\qquad \ a\in \{{\cal K}^{(\infty)}_0,{\cal K}^{(\infty)}_1, {\cal Z}^{(\infty)}_1, \tilde{\cal Z}^{(\infty)}_1\}\ . \label{symdiag}
\eeqa  
According to these results, we then conclude that the $q-$Onsager and augmented $q-$Onsager algebras with defining relations (\ref{Talg}) and (\ref{Taug}) emerge as the non-Abelian symmetry of the Hamiltonian (\ref{Hsemi}) for generic {\it non-diagonal} and {\it diagonal} boundary conditions ($k_\pm=0$), respectively. Despite of the fact that this property played no role in previous analysis, it has not been observed previously in the literature, to our knowledge.\vspace{1mm}

Besides, we would like to make a few comments. In \cite{PS}, recall that common algebraic structures were exhibited between certain finite lattice models and conformal field theories. For instance, it was shown that the Hamiltonian of the XXZ open spin chain can be understood as the discrete analog of the Virasoro generator $L_0$.
In this picture, the Temperley-Lieb algebra and the Virasoro algebra share similar properties. For instance, consider the Hamiltonians
\beqa
\qquad H^{(\pm)}_{\frac{1}{2}XXZ}&=&-\frac{1}{2}\sum_{k=1}^{\infty}\Big(\sigma_1^{k+1}\sigma_1^{k}+\sigma_2^{k+1}\sigma_2^{k} + \Delta\sigma_3^{k+1}\sigma_3^{k}\Big) \pm \frac{(q-q^{-1})}{4}\sigma^1_3 \ 
\label{HsemiPS}\  
\eeqa
which can be obtained as the thermodynamic limit of the $U_q(sl_2)-$symmetric XXZ open spin chain. As described in \cite{PS}, the special value  of the boundary field (compared with (\ref{Hsemi}))  plays a very singular role: for the deformation parameter $q=\exp(i\pi/\mu(\mu+1))$, $\mu\notin {\mathbb Q}$, the central charge of the Virasoro algebra associated with $H^{(-)}_{\frac{1}{2}XXZ}$ was identified with $c=1-6/\mu(\mu+1)$
and the Hamiltonian's spectrum was expressed in terms of conformal dimensions. In light of previous results, for the special class of diagonal boundary conditions $(+)$ (resp. $(-)$) associated with $\epsilon_+=0,\epsilon_-\neq 0$ (resp. $\epsilon_-=0,\epsilon_+\neq 0$) some remarkable properties are then expected. Indeed, according to the analysis above,  the vacuum vector of the Hamiltonian $H^{(+)}_{\frac{1}{2}XXZ}$ (resp. $ H^{(-)}_{\frac{1}{2}XXZ}$) is given by $|B_+\rangle$ (resp.  $|B_-\rangle$). In other words, the spectrum of the Hamiltonians (\ref{HsemiPS}) is classified according to the eigenvalues of the fundamental generators of the augmented $q-$Onsager algebra. Moreover, it is worth mentioning that the realizations of the $O_q(\widehat{sl_2})$ currents in terms of type II $q-$vertex operators such as (\ref{realW}) share some analogy with currents arising in the study of the $q-$deformed Virasoro algebra \cite{LP} or currents exihibited in the context of conformal field theory \cite{Kau}. In view of this, a relation between the representation theory of $O_q(\widehat{sl_2})$ and the $q-$Virasoro algebra may be investigated. 
\vspace{1mm}

Finally, let us mention that the formulation (\ref{transf}) can be applied to other integrable models directly, for instance the XXZ open chain with higher spins or alternating spins. In these cases, the results of \cite{Idz} have to be considered. More generally, it can be extended to models with higher symmetries (see e.g. \cite{BeR}) in which case generalizations of the current algebra (\ref{ec1})-(\ref{ec16}) are needed. A first step in this direction has been passed in \cite{BB1,Kolb}, where generalizations of the $q-$Onsager algebra and twisted $q-$ Yangians \cite{MRS} have been proposed (for some applications, see \cite{BF}). In this picture, the problem of the diagonalization of the transfer matrix generalizing (\ref{transf}) with (\ref{Iinf}) relies on a better understanding of the representation theory associated with certain coideal subalgebras of $U_q(\widehat{g})$. Another interesting direction, obviously inspired by the conformal field theory program, concerns the family of $q-$difference equations for the correlation functions in Onsager's picture. We intend to discuss some of these problems elsewhere.
\vspace{3mm} 

\noindent{\bf Acknowledgements:}  
P.B. thanks T. Kojima for discussions. S.B. thanks N. Cramp\'e for discussions and also the LMPT for hospitality, where part of this work has been done.  
\vspace{1cm}

\newpage

\appendix

\section{Generators of the infinite dimensional algebras ${\cal A}_q$ and ${\cal A}^{diag}_q$}
\vspace{2mm}

$\bullet$ {\bf Elements generating ${\cal A}_q$:}  For generic values of the parameters $\epsilon_\pm$, $k_\pm\neq 0$ and $N\in{\mathbb N}$, the elements ${\textsf W}_{-k},{\textsf W}_{k+1}, {\textsf G}_{k+1},{\tilde{\textsf G}}_{k+1}$ ($4N$ in total) act on $N-$tensor product (evaluation) representations of $U_q(\widehat{sl_2})$, and depend solely on the $N-$parameters $v_k$ and spin$-j_k$ for $k=1,...,N$. Define $w_0^{(j_k)}=q^{2j_k+1}+q^{-2j_k-1}$.  According to the ordering of the vector spaces 
\beqa
{\cal V^{(N)}}= {\cal V}_N \otimes \cdot \cdot\cdot \otimes {\cal V}_2 \otimes {\cal V}_1\ ,
\eeqa
they act as \cite{BK1} (see also \cite{BK}):
\beqa
\qquad \qquad {\textsf W}_{-k}^{(N)}&=&\frac{(w_0^{(j_{N})}-(q+q^{-1})q^{2s_3})}{(q+q^{-1})}\otimes
{\textsf W}_{k}^{(N-1)}
-\frac{(v_{N}^2+v_{N}^{-2})}{(q+q^{-1})}I\!\!I\otimes {\textsf W}_{-k+1}^{(N-1)} +\ \frac{(v_N^2+v_N^{-2})w_0^{(j_N)}}{(q+q^{-1})^2}{\textsf W}_{-k+1}^{(N)}
\label{rep0}\\
&&\ \ \ + \ \frac{(q-q^{-1})}{k_+k_-(q+q^{-1})^2}
\left(k_+v_Nq^{1/2}S_+q^{s_3}\otimes
{\textsf G}_{k}^{(N-1)}+k_-v_N^{-1}q^{-1/2}S_-q^{s_3}\otimes {\tilde {\textsf G}}_{k}^{(N-1)}\right)\nonumber\\
&&\ \ \ +\ q^{2s_3}\otimes {\textsf W}_{-k}^{(N-1)}
\ ,\nonumber\\
{\textsf W}_{k+1}^{(N)}&=&\frac{(w_0^{(j_{N})}-(q+q^{-1})q^{-2s_3})}{(q+q^{-1})}\otimes
{\textsf W}_{-k+1}^{(N-1)}
-\frac{(v_{N}^2+v_{N}^{-2})}{(q+q^{-1})}I\!\!I\otimes {\textsf W}_{k}^{(N-1)} +\ \frac{(v_N^2+v_N^{-2})w_0^{(j_N)}}{(q+q^{-1})^2}{\textsf W}_{k}^{(N)}
\nonumber\\
&&\ \ \ +\ \frac{(q-q^{-1})}{k_+k_-(q+q^{-1})^2}
\left(k_+v^{-1}_Nq^{-1/2}S_+q^{-s_3}\otimes
{\textsf G}_{k}^{(N-1)}+k_-v_Nq^{1/2}S_-q^{-s_3}\otimes {\tilde {\textsf G}}_{k}^{(N-1)}\right)\nonumber\\
&&\ \ \ +\ q^{-2s_3}\otimes {\textsf W}_{k+1}^{(N-1)}
\ ,\nonumber\\
{\textsf G}_{k+1}^{(N)}&=& 
\frac{k_-(q-q^{-1})^2}{k_+(q+q^{-1})}
S_-^2\otimes {\tilde {\textsf G}}_{k}^{(N-1)}
-\frac{1}{(q+q^{-1})}(v_N^{2}q^{2s_3}+v_N^{-2}q^{-2s_3})\otimes {\textsf G}_{k}^{(N-1)} 
+I\!\!I \otimes {\textsf G}_{k+1}^{(N-1)}\nonumber\\
&& + (q-q^{-1})\left(
k_-v_Nq^{-1/2}S_-q^{s_3}\otimes \big({\textsf W}_{-k}^{(N-1)}-{\textsf W}_{k}^{(N-1)}\big)
+k_-v_N^{-1}q^{1/2}S_-q^{-s_3}\otimes \big({\textsf W}_{k+1}^{(N-1)}-{\textsf W}_{-k+1}^{(N-1)}\big)
\right)\nonumber\\
&&+\frac{(v_N^2+v_N^{-2})w_0^{(j_N)}}{(q+q^{-1})^2}{\textsf G}_{k}^{(N)}\ ,\nonumber\\
\nonumber
{\tilde{\textsf G}}_{k+1}^{(N)}&=& 
\frac{k_+(q-q^{-1})^2}{k_-(q+q^{-1})}
S_+^2\otimes {{\textsf G}}_{k}^{(N-1)}
-\frac{1}{(q+q^{-1})}(v_N^{2}q^{-2s_3}+v_N^{-2}q^{2s_3})\otimes {\tilde{\textsf G}}_{k}^{(N-1)} 
+I\!\!I \otimes {\tilde{\textsf G}}_{k+1}^{(N-1)}\nonumber\\
&& + (q-q^{-1})\left(
k_+v^{-1}_Nq^{1/2}S_+q^{s_3}\otimes \big({\textsf W}_{-k}^{(N-1)}-{\textsf W}_{k}^{(N-1)}\big)
+k_+v_Nq^{-1/2}S_+q^{-s_3}\otimes \big({\textsf W}_{k+1}^{(N-1)}-{\textsf W}_{-k+1}^{(N-1)}\big)
\right)\nonumber\\
&&+\frac{(v_N^2+v_N^{-2})w_0^{(j_N)}}{(q+q^{-1})^2}{\tilde{\textsf G}}_{k}^{(N)}\ ,\nonumber
\eeqa
where, for the special case $k=0$ we identify\,\footnote{Although the notation is ambiguous, the reader must keep in mind that ${{\textsf W}}_{k}^{(N)}|_{k=0}\neq {{\textsf W}}_{-k}^{(N)}|_{k=0}$\ ,${{\textsf W}}_{-k+1}^{(N)}|_{k=0}\neq {{\textsf W}}_{k+1}^{(N)}|_{k=0}$\ for any $N$.}
\beqa
{{\textsf W}}_{k}^{(N)}|_{k=0}\equiv 0\ ,\quad {{\textsf W}}_{-k+1}^{(N)}|_{k=0}\equiv 0\ ,\quad {\textsf G}_{k}^{(N)}|_{k=0}={\tilde{\textsf G}}_{k}^{(N)}|_{k=0}\equiv \frac{k_+k_-(q+q^{-1})^2}{(q-q^{-1})}I\!\!I^{(N)}\ .\label{not0}
\eeqa
In addition, one has the ``initial'' $c-$number conditions
\beqa
{{\textsf W}}_{0}^{(0)}\equiv \epsilon^{(0)}_+\ ,\quad {{\textsf W}}_{1}^{(0)}\equiv \epsilon^{(0)}_-\qquad  \mbox{and}\qquad
{\textsf G}_{1}^{(0)}={\tilde{\textsf G}}_{1}^{(0)}\equiv \epsilon^{(0)}_+\epsilon^{(0)}_-(q-q^{-1})\ .\label{initrep0}
\eeqa
\vspace{2mm}

$\bullet$ {\bf Elements generating ${\cal A}^{diag}_q$:} All expressions below are derived from above expressions, through the substitutions:
\beqa
&&{\textsf W}_{-k}^{(N)}\rightarrow {\textsf K}_{-k}^{(N)}\ ,\quad {\textsf W}_{k+1}^{(N)}\rightarrow {\textsf K}_{k+1}^{(N)}\ ,\label{subsKZ}\\
&&{\textsf G}_{k+1}^{(N)}\rightarrow k_-\big({\textsf Z}_{k+1}^{(N)} + \epsilon_+\epsilon_-(q-q^{-1})\ I\!\!I^{(N)} \big)\ ,\quad {\tilde {\textsf G}}_{k+1}^{(N)}\rightarrow k_+\big({\tilde {\textsf Z}}_{k+1}^{(N)}+ \epsilon_+\epsilon_-(q-q^{-1})\ I\!\!I^{(N)} \big)\ \nonumber
\eeqa
and then setting $k_\pm=0$. It yields to:
\beqa
\qquad \qquad {\textsf K}_{-k}^{(N)}&=&\frac{(w_0^{(j_{N})}-(q+q^{-1})q^{2s_3})}{(q+q^{-1})}\otimes
{\textsf K}_{k}^{(N-1)}
-\frac{(v_{N}^2+v_{N}^{-2})}{(q+q^{-1})}I\!\!I\otimes {\textsf K}_{-k+1}^{(N-1)} +\ \frac{(v_N^2+v_N^{-2})w_0^{(j_N)}}{(q+q^{-1})^2}{\textsf K}_{-k+1}^{(N)}
\label{rep}\\
&&\ \ \ + \ \frac{(q-q^{-1})}{(q+q^{-1})^2}
\left(v_Nq^{1/2}S_+q^{s_3}\otimes
{\textsf Z}_{k}^{(N-1)}+v_N^{-1}q^{-1/2}S_-q^{s_3}\otimes {\tilde {\textsf Z}}_{k}^{(N-1)}\right)\ +\ q^{2s_3}\otimes {\textsf K}_{-k}^{(N-1)}
\ ,\nonumber\\
{\textsf K}_{k+1}^{(N)}&=&\frac{(w_0^{(j_{N})}-(q+q^{-1})q^{-2s_3})}{(q+q^{-1})}\otimes
{\textsf K}_{-k+1}^{(N-1)}
-\frac{(v_{N}^2+v_{N}^{-2})}{(q+q^{-1})}I\!\!I\otimes {\textsf K}_{k}^{(N-1)} +\ \frac{(v_N^2+v_N^{-2})w_0^{(j_N)}}{(q+q^{-1})^2}{\textsf K}_{k}^{(N)}
\nonumber\\
&&\ \ \ +\ \frac{(q-q^{-1})}{(q+q^{-1})^2}
\left(v^{-1}_Nq^{-1/2}S_+q^{-s_3}\otimes
{\textsf Z}_{k}^{(N-1)}+v_Nq^{1/2}S_-q^{-s_3}\otimes {\tilde {\textsf Z}}_{k}^{(N-1)}\right)\ +\ q^{-2s_3}\otimes {\textsf K}_{k+1}^{(N-1)}
\ ,\nonumber\\
{\textsf Z}_{k+1}^{(N)}&=& 
\frac{(q-q^{-1})^2}{(q+q^{-1})}
S_-^2\otimes {\tilde {\textsf Z}}_{k}^{(N-1)}
-\frac{1}{(q+q^{-1})}(v_N^{2}q^{2s_3}+v_N^{-2}q^{-2s_3})\otimes {\textsf Z}_{k}^{(N-1)} 
+I\!\!I \otimes {\textsf Z}_{k+1}^{(N-1)}\nonumber\\
&& + (q-q^{-1})\left(
v_Nq^{-1/2}S_-q^{s_3}\otimes \big({\textsf K}_{-k}^{(N-1)}-{\textsf K}_{k}^{(N-1)}\big)
+v_N^{-1}q^{1/2}S_-q^{-s_3}\otimes \big({\textsf K}_{k+1}^{(N-1)}-{\textsf K}_{-k+1}^{(N-1)}\big)
\right)\nonumber\\
&&+\frac{(v_N^2+v_N^{-2})w_0^{(j_N)}}{(q+q^{-1})^2}{\textsf Z}_{k}^{(N)}\ ,\nonumber\\
\nonumber\\
{\tilde{\textsf Z}}_{k+1}^{(N)}&=& 
\frac{(q-q^{-1})^2}{(q+q^{-1})}
S_+^2\otimes {{\textsf Z}}_{k}^{(N-1)}
-\frac{1}{(q+q^{-1})}(v_N^{2}q^{-2s_3}+v_N^{-2}q^{2s_3})\otimes {\tilde{\textsf Z}}_{k}^{(N-1)} 
+I\!\!I \otimes {\tilde{\textsf Z}}_{k+1}^{(N-1)}\nonumber\\
&& + (q-q^{-1})\left(
v^{-1}_Nq^{1/2}S_+q^{s_3}\otimes \big({\textsf K}_{-k}^{(N-1)}-{\textsf K}_{k}^{(N-1)}\big)
+v_Nq^{-1/2}S_+q^{-s_3}\otimes \big({\textsf K}_{k+1}^{(N-1)}-{\textsf K}_{-k+1}^{(N-1)}\big)
\right)\nonumber\\
&&+\frac{(v_N^2+v_N^{-2})w_0^{(j_N)}}{(q+q^{-1})^2}{\tilde{\textsf Z}}_{k}^{(N)}\ .\nonumber
\eeqa
As before, we identify\,\footnote{Remind that ${{\textsf K}}_{k}^{(N)}|_{k=0}\neq {{\textsf K}}_{-k}^{(N)}|_{k=0}$\ ,${{\textsf K}}_{-k+1}^{(N)}|_{k=0}\neq {{\textsf K}}_{k+1}^{(N)}|_{k=0}$\ for any $N$.}
\beqa
{{\textsf K}}_{k}^{(N)}|_{k=0}\equiv 0\ ,\quad {{\textsf K}}_{-k+1}^{(N)}|_{k=0}\equiv 0\ ,\quad {\textsf Z}_{k}^{(N)}|_{k=0}\equiv 0\ ,\quad {\tilde{\textsf Z}}_{k}^{(N)}|_{k=0}\equiv 0\ \label{not}
\eeqa
together with the ``initial'' $c-$number conditions
\beqa
{{\textsf K}}_{0}^{(0)}\equiv \epsilon_+\ ,\quad {{\textsf K}}_{1}^{(0)}\equiv \epsilon_-\qquad  \mbox{and}\qquad
{\textsf Z}_{1}^{(0)}={\tilde{\textsf Z}}_{1}^{(0)}\equiv 0\ .\label{initrep}
\eeqa
\vspace{2mm}

$\bullet$ {\bf Application to the homogeneous XXZ open spin-$\frac{1}{2}$ chain:} For generic non-diagonal $k_\pm\neq 0$ or diagonal $k_\pm\equiv 0$ boundary conditions, the generators of the infinite dimensional algebras ${\cal A}_q$ or  ${\cal A}^{diag}_q$ are simply given, respectively, by:
\beqa
{\cal W}^{(N)}_{-l}&\equiv&(\otimes_{k=1}^{N}\pi^{(\frac{1}{2})})[{\textsf W}_{-l}^{(N)}]|_{v_k=1}\ ,\qquad
{\cal W}^{(N)}_{l+1}\equiv(\otimes_{k=1}^{N}\pi^{(\frac{1}{2})})[{\textsf W}_{l+1}^{(N)}]|_{v_k=1}\ ,\nonumber\\
{\cal G}^{(N)}_{l+1}&\equiv&(\otimes_{k=1}^{N}\pi^{(\frac{1}{2})})[{\textsf G}_{l+1}^{(N)}]|_{v_k=1}\ ,\qquad
\ \  {\tilde{\cal G}}^{(N)}_{l+1}\equiv(\otimes_{k=1}^{N}\pi^{(\frac{1}{2})})[{\tilde{\textsf G}}_{l+1}^{(N)}]|_{v_k=1}\ \nonumber
\eeqa
or
\beqa
{\cal K}^{(N)}_{-l}&\equiv&(\otimes_{k=1}^{N}\pi^{(\frac{1}{2})})[{\textsf K}_{-l}^{(N)}]|_{v_k=1}\ ,\qquad
 {\cal K}^{(N)}_{l+1}\equiv(\otimes_{k=1}^{N}\pi^{(\frac{1}{2})})[{\textsf K}_{l+1}^{(N)}]|_{v_k=1}\ ,\nonumber\\
 {\cal Z}^{(N)}_{l+1}&\equiv&(\otimes_{k=1}^{N}\pi^{(\frac{1}{2})})[{\textsf Z}_{l+1}^{(N)}]|_{v_k=1}\ ,\qquad
\ \  {\tilde{\cal Z}}^{(N)}_{l+1}\equiv(\otimes_{k=1}^{N}\pi^{(\frac{1}{2})})[{\tilde{\textsf Z}}_{l+1}^{(N)}]|_{v_k=1}\ \label{opXXZ}  
\eeqa
for $l\in 0,...,N-1$. Here we considered the two-dimensional representation  $\pi^{(1/2)}$ given by:
\beqa
\pi^{(1/2)}[S_\pm]=\sigma_\pm \qquad \mbox{and}\qquad \pi^{(1/2)}[s_3]=\sigma_3/2\ .\label{2dim}
\eeqa
\vspace{1cm}

\section{Drinfeld-Jimbo presentation of $U_q(\widehat{sl_2})$}
\vspace{2mm}
Define the extended Cartan matrix $\{a_{ij}\}$ ($a_{ii}=2$,\ $a_{ij}=-2$ for $i\neq j$). The quantum affine algebra $U_{q}(\widehat{sl_2})$ is generated by the elements $\{h_j,e_j,f_j\}$, $j\in \{0,1\}$ which satisfy the defining relations
\beqa [h_i,h_j]=0\ , \quad [h_i,e_j]=a_{ij}e_j\ , \quad
[h_i,f_j]=-a_{ij}f_j\ ,\quad
[e_i,f_j]=\delta_{ij}\frac{q^{h_i}-q^{-h_i}}{q-q^{-1}}\
\nonumber\eeqa
together with the $q-$Serre relations
\beqa [e_i,[e_i,[e_i,e_j]_{q}]_{q^{-1}}]=0\ ,\quad \mbox{and}\quad
[f_i,[f_i,[f_i,f_j]_{q}]_{q^{-1}}]=0\ . \label{defUq}\eeqa
The sum ${\it C}=h_0+h_1$ is the central element of the algebra. The
Hopf algebra structure is ensured by the existence of a
comultiplication $\Delta: U_{q}(\widehat{sl_2})\mapsto U_{q}(\widehat{sl_2})\otimes U_{q}(\widehat{sl_2})$, antipode ${\cal S}: U_{q}(\widehat{sl_2})\mapsto U_{q}(\widehat{sl_2})$ 
and a counit ${\cal E}: U_{q}(\widehat{sl_2})\mapsto {\mathbb C}$ with
\beqa \Delta(e_i)&=&e_i\otimes 1 +
q^{h_i}\otimes e_i\ ,\nonumber \\
 \Delta(f_i)&=&f_i\otimes q^{-h_i} + 1 \otimes f_i\ ,\nonumber\\
 \Delta(h_i)&=&h_i\otimes 1 + 1 \otimes h_i\ ,\label{coprod}
\eeqa
\beqa {\cal S}(e_i)=-q^{-h_i}e_i\ ,\quad {\cal S}(f_i)=-f_iq^{h_i}\ ,\quad {\cal S}(h_i)=-h_i \qquad {\cal S}({1})=1\
\label{antipode}\nonumber\eeqa
and\vspace{-0.3cm}
\beqa {\cal E}(e_i)={\cal E}(f_i)={\cal
E}(h_i)=0\ ,\qquad {\cal E}(1)=1\
.\label{counit}\nonumber\eeqa
Note that the opposite coproduct $\Delta'$ can be similarly defined with $\Delta'\equiv \sigma
\circ\Delta$ where the permutation map $\sigma(x\otimes y
)=y\otimes x$ for all $x,y\in U_{q}(\widehat{sl_2})$ is used.\vspace{2mm} 

The (evaluation in the principal gradation) endomorphism  $\pi_\zeta: U_{q}(\widehat{sl_2}) \mapsto \mathrm{End}({\cal V}_\zeta)$ is chosen such that $({\cal V}\equiv{\mathbb C}^2)$
\beqa 
&&\pi_\zeta[e_1]= \zeta\sigma_+\ , \qquad \ \ \ \ \ \pi_\zeta[e_0]= \zeta\sigma_-\ , \nonumber\\
&&\pi_\zeta[f_1]=
\zeta^{-1}\sigma_-\ ,\qquad \pi_\zeta[f_0]= \zeta^{-1}\sigma_+\ ,\nonumber\\
\ &&\pi_\zeta[q^{h_1}]= q^{\sigma_3}\ ,\quad \qquad  \ \pi_\zeta[q^{h_0}]=
q^{-\sigma_3}\ ,\ \label{evalrep}
\eeqa
in terms of the Pauli matrices $\sigma_\pm,\sigma_3$:
\beqa
\sigma_+=\left(
\begin{array}{cc}
 0    & 1 \\
 0 & 0 
\end{array} \right) \ ,\qquad
\sigma_- =\left(
\begin{array}{cc}
 0    & 0 \\
 1 & 0 
\end{array} \right) \ ,\qquad
\sigma_3 =\left(
\begin{array}{cc}
 1    & 0 \\
 0 & -1 
\end{array} \right) \ .\label{Pauli}\nonumber
\eeqa

The $R-$matrix here considered is the solution of the intertwining equation:
\beqa
R(\zeta_1/\zeta_2) (\pi_{\zeta_1} \otimes \pi_{\zeta_2})\, \Delta(x)= (\pi_{\zeta_1} \otimes \pi_{\zeta_2})\, (\sigma \circ \Delta(x)) R(\zeta_1/\zeta_2) \  .\nonumber
\eeqa
According to above definitions and up to an overall scalar factor, in the principal picture it reads:
\begin{align}
R(\zeta) = \frac{1}{\kappa(\zeta)}\left(
\begin{array}{cccc} 
1    & 0 & 0 & 0 \\
0  & \frac{(1 -  \zeta^{2})q}{1 -  q^2\zeta^2} & \frac{(1-q^2)\zeta}{1 -  q^2\zeta^2} & 0 \\
0  &  \frac{(1-q^2)\zeta}{1 -  q^2\zeta^2}  &  \frac{(1 -  \zeta^{2})q}{1 -  q^2\zeta^2}  &  0 \\
0 & 0 & 0 & 1
\end{array} \right) \ .\label{R}
\end{align}
where the scalar factor
\beqa
\kappa(\zeta)=\zeta \frac{(q^4\zeta^2;q^4)_\infty  (q^2\zeta^{-2};q^4)_\infty}{(q^4\zeta^{-2};q^4)_\infty (q^2\zeta^2;q^4)_\infty}\ , \qquad (z;p)_\infty=\prod_{n=0}^{\infty} (1-zp^n)
\eeqa
is chosen to ensure unitarity and crossing symmetry of the $R-$matrix:
\beqa
R(\zeta)R(\zeta^{-1})&=&  I\!\!I \otimes  I\!\!I \ ,\label{unit} \\
R^{\epsilon'_2 \epsilon_1}_{\epsilon_2 \epsilon'_1}(\zeta^{-1}) &=& R^{-\epsilon'_1 \epsilon'_2}_{-\epsilon_1 \epsilon_2}(-q^{-1}\zeta) \ .\nonumber
\eeqa
\vspace{1mm}

\newpage

\section{The $q-$Vertex operators of $U_q(\widehat{sl_2})$}
\vspace{2mm}
The so-called type I and type II $q-$vertex operators satisfy the commutation relations:
\begin{eqnarray}
&&
\Phi_{\epsilon_2}(\zeta_2)\Phi_{\epsilon_1}(\zeta_1)=
\sum_{\epsilon_1',\epsilon_2'}
R^{\epsilon_1'\,\epsilon_2'}_{\epsilon_1\,\epsilon_2}(\zeta_1/\zeta_2)
\Phi_{\epsilon_1'}(\zeta_1)\Phi_{\epsilon_2'}(\zeta_2)\ ,
\label{eqn:comI}\\
&&\Psi_{\mu_1'}^*(\zeta_1)\Psi_{\mu_2'}^*(\zeta_2)
=-\sum_{\mu_1,\mu_2}R^{\mu_1'\,\mu_2'}_{\mu_1\,\mu_2}(\zeta_1/\zeta_2)
\Psi_{\mu_2}^*(\zeta_2)\Psi_{\mu_1}^*(\zeta_1)\ ,
\label{eqn:psicom}\\
&&\Phi_{\epsilon}(\zeta_1)\Psi_{\mu}^*(\zeta_2)=
\tau(\zeta_1/\zeta_2)\Psi_{\mu}^*(\zeta_2)\Phi_{\epsilon}(\zeta_1)\ .
\label{eqn:phipsi}
\end{eqnarray}
Here
\begin{equation}\label{eqn:tau}
\tau(\zeta)=
\zeta^{-1}{\Theta_{q^4}(q\zeta^{2}) \over \Theta_{q^4}(q\zeta^{-2})}\ ,
\qquad
\Theta_p(z)=(z;p)_\infty(pz^{-1};p)_\infty (p;p)_\infty\ .\nonumber
\end{equation}

Define $\Phi^*_{\epsilon}(\zeta)=\Phi_{-\epsilon}(-q^{-1}\zeta)$. The type I vertex operators satisfy the invertibility relations
\begin{equation}\label{eqn:invert}
g\,\sum_\epsilon \Phi^*_\epsilon(\zeta)\Phi_\epsilon(\zeta)=\id\ ,
\qquad
g\,\Phi_{\epsilon_1}(\zeta)\Phi^*_{\epsilon_2}(\zeta)=\delta_{\epsilon_1\epsilon_2}\id  \quad \mbox{with}\quad g=\frac{\pr{q^2;q^4}}{\pr{q^4;q^4}}\ .
\end{equation}

Type I and type II $q-$vertex operators admit a bosonic realization \cite{vertex}.
\ For $i=0,1$, consider the bosonic Fock space
\[
\H^{(i)}=\C[a_{-1},a_{-2},\cdots]
\otimes \left(\oplus_{n\in\Z}\C e^{\Lambda_i+n\alpha}\right)
\]
where the commutation relations of $a_n$ are given by
\[
[a_m,a_n]=\delta_{m+n,0}{[m][2m]\over m}~, \quad\mbox{with}\quad
\quad m,n\neq 0 \quad\mbox{and}\qquad [n]=\frac{q^n-q^{-n}}{q-q^{-1}}\ .
\]
Define $[\partial,\alpha]=2$, $[\partial, \Lambda_0]=0$
and $\Lambda_1=\Lambda_0+\alpha/2$. The highest weight vector of $\H^{(i)}$ is given by
$|{i}\rangle=1\otimes e^{\Lambda_i}$ and the operators $e^\beta$, $z^\partial$ act as
\[
e^\beta . e^\gamma=e^{\beta+\gamma},
\qquad
z^{\partial}.e^\gamma=z^{[\partial,\gamma]}e^\gamma \ .\]

\vspace{1mm}
The bosonic realization for the type I and type II $q-$vertex operators reads \cite{vertex}:
\begin{eqnarray}
\Phi^{(1-i,i)}_-(\zeta)
&=&
e^{P(\zeta^2)}e^{Q(\zeta^2)}\otimes
e^{\alpha/2}(-q^3\zeta^2)^{(\partial +i)/2}\zeta^{-i}\ ,
\label{eqn:phim}\\
\Phi^{(1-i,i)}_+(\zeta)&=&
\oint_{C_1} {dw \over 2\pi i}
{(1-q^2)w\zeta \over q(w-q^2\zeta^2)(w-q^4\zeta^2)}:\Phi^{(1-i,i)}_-(\zeta)X^-(w):\ ,
\nonumber\\
&&\label{eqn:phip}\\
\Psi^{*(1-i,i)}_-(\zeta)&=&
e^{-P(q^{-1}\zeta^2)}e^{-Q(q\zeta^2)}\otimes
e^{-\alpha/2}(-q^3\zeta^2)^{(-\partial +i)/2}\zeta^{1-i}\ ,
\label{eqn:psim}\\
\Psi^{*(1-i,i)}_+(\zeta)&=&
\oint_{C_2} {dw \over 2\pi i}
{q^2(1-q^2)\zeta \over (w-q^2\zeta^2)(w-q^4\zeta^2)}:\Psi^{*(1-i,i)}_-(\zeta)X^+(w):\ ,
\label{eqn:psip}
\end{eqnarray}
where
\begin{eqnarray}
X^{\pm}(z)&=&e^{R^\pm(z)}e^{S^\pm(z)}\otimes e^{\pm \alpha}z^{\pm\partial}\ ,
\label{eqn:xpm}
\end{eqnarray}

\begin{eqnarray}
&&P(z)=\sum_{n=1}^\infty {\frac{a_{-n}}{[2n]}}q^{7n/2}z^n\ ,
\qquad
Q(z)=-\sum_{n=1}^\infty {\frac{a_{n}}{[2n]}}q^{-5n/2}z^{-n}\ ,
\label{eqn:PQ} \nonumber\\
&&R^{\pm}(z)=
\pm \sum_{n=1}^\infty {\frac{a_{-n}}{[n]}}q^{\mp n/2}z^n\ ,
\quad
S^\pm(z)=\mp \sum_{n=1}^\infty {\frac{a_{n}}{[n]}}q^{\mp n/2}z^{-n}\ .
\label{eqn:RS}\nonumber
\end{eqnarray}
The integration contours encircle $w=0$ in such a way that
\begin{eqnarray*}
C_1&:&\hbox{ $q^4\zeta^2$ is inside and $q^2\zeta^2$ is outside}, \\
C_2&:&\hbox{ $q^4\zeta^2$ is outside and $q^2\zeta^2$ is inside}.
\end{eqnarray*}

\end{document}